\documentclass[a4paper,fleqn,usenatbib]{mnras}
\usepackage{newtxtext,newtxmath}

\usepackage[T1]{fontenc}
\usepackage{ae,aecompl}

\usepackage{graphicx}	
\usepackage{amsmath}	
\usepackage{amssymb}	

\def\Planck{\textit{Planck}}

\title[Directionality in the \textit{Planck} data]{Testing for directionality in the
\textit{Planck} polarization and lensing data}

\author[M. Ghrear et al.]{
Majd Ghrear$^{1}$\thanks{E-mail: majd.ghrear@alumni.ubc.ca},
Emory F. Bunn$^{2}$,
Dagoberto Contreras$^{1}$ and
Douglas Scott$^{1}$
\\
$^{1}$Department of Physics and Astronomy, University of British Columbia, Vancouver, BC V6T 1Z1, Canada\\
$^{2}$Department of Physics, University of Richmond, Richmond, VA 23173, USA\\
}

\date{Accepted XXX. Received YYY; in original form ZZZ}

\pubyear{2018}

\begin{document}
\label{firstpage}
\pagerange{\pageref{firstpage}--\pageref{lastpage}}
\maketitle

\begin{abstract}
In order to better analyse the polarization of the cosmic microwave background
(CMB), which is dominated by emission from our Galaxy,
we need tools that can detect residual foregrounds in cleaned CMB maps.
Galactic foregrounds introduce statistical anisotropy and directionality to
the polarization pseudo-vectors of the CMB, which can be investigated by using
the $\mathcal{D}$ statistic of Bunn and Scott. This statistic is rapidly
computable and capable of investigating a broad range of data products for
directionality. We demonstrate the application of this statistic to detecting
foregrounds in polarization maps by analysing the uncleaned \Planck\
\textit{2018} frequency maps.  For the \Planck\ \textit{2018} CMB maps, we find no evidence for residual foreground contamination. In order to examine
the sensitivity of the $\mathcal{D}$ statistic, we add a varying fraction of
the polarized thermal dust and synchrotron foreground maps to the CMB maps and
show the percent-level foreground contamination that would be detected
with 95 percent confidence. We also demonstrate application of the
$\mathcal{D}$ statistic to another data product by analysing the gradient of
the minimum-variance CMB lensing potential map (i.e., the deflection angle)
for directionality. We find no excess directionality in the lensing potential
map when compared to the simulations provided by the Planck Collaboration.
\end{abstract}




\section{Introduction}
\label{sec:intro}

Anisotropies in the cosmic microwave background (CMB) provide a means of
probing the large-scale structure of the Universe. Analysing the polarization of the CMB anisotropies provides a wealth of cosmological information in addition to that available from the temperature anisotropies. One exciting possibility is the chance to detect primordial gravitational waves through the measurement of \textit{B}-mode polarization \citep{1997NewA....2..323H, 1997PhRvD..55.7368K, 1997PhRvL..78.2054S, 1998PhRvD..57..685K}.

Unfortunately, the magnitude of the CMB polarization anisotropies is small compared to those of temperature, small enough that the primordial signal is dominated by foreground emission. Specifically, synchrotron and dust emission from our Galaxy contaminate uncleaned polarization maps, and it is important to test whether cleaned maps are indeed free of foregrounds (and other systemic effects).

This paper focuses on using a test for directionality as a proxy for Galactic foreground contamination and other systematic effects. The cosmological principle implies that the CMB is statistically isotropic, whereas foregrounds produced by the Galaxy have a preferred axis.
The $\mathcal{D}$ statistic of \cite{2000MNRAS.313..331B} \citep[see also][]{2007MNRAS.381....2H} provides a measure of global directionality of a map in a general way and is an effective test for a broad range of types of anisotropic residuals. Here, we apply the $\mathcal{D}$ statistic to test for foreground contamination in various polarization maps provided by the Planck Collaboration. We examine the sensitivity of this statistic and show that it is well suited for detection of foreground contamination, since both synchrotron and dust emission have strong directionality on large scales.

The $\mathcal{D}$ statistic has a simple interpretation as a measure of directionality and is extremely rapid to compute. In Section~\ref{sec:dstat} we introduce and define the $\mathcal{D}$ statistic, in Sections~\ref{sec:freq}--\ref{sec:foreground_contamination} we show our results of the $\mathcal{D}$ statistic as applied to foreground maps, raw frequency maps and CMB maps. In Section~\ref{sec:lensing} we additionally perform a general analysis of the directionality in the \Planck\ lensing maps. Finally, we conclude in Section~\ref{sec:conclusions}.

\section{The $\mathcal{D}$ statistic}
\label{sec:dstat}

One can imagine defining many different statistical approaches for deciding if all-sky data have a preferred direction. Some of these have been motivated, for example, by searches for axial symmetry, such as might be expected in some topologically small universe scenarios \citep[e.g.,][]{1986PhLA..115...97E,1993PhRvL..71...20S,1996ApJ...468..457D}. Other approaches
attempt to be more agnostic about the form that the directionality might take. The $\mathcal{D}$ statistic, presented by \cite{2000MNRAS.313..331B}, falls in this latter category. It is defined over a pixelized map as
\begin{equation}
    \mathcal{D} \equiv \frac{\max_{\hat{\mathbfit{n}}}{f(\hat{\mathbfit {n}})}}{\min_{\hat{\mathbfit {n}}}{f(\hat{\mathbfit {n}})}},
    \label{D}
\end{equation}
where the vector $\hat{\mathbfit{n}}$ ranges over the celestial sphere and
$f(\hat{\mathbfit{n}})$ is defined as
\begin{equation}
    f(\hat{\mathbfit{n}}) \equiv \sum_{p=1}^{N} w_p({\hat{\mathbfit{n}} \cdot \mathbfit{g}_p})^2.
  \label{f}
\end{equation}
Here, the sum, $\sum_{p=1}^{N}$, is over all unmasked pixels. The weights, $w_p$, are chosen to remove the effects of noise structure and masking of the sky. A local vector, $\mathbfit{g}_p$, is assigned to each pixel and $f(\hat{\mathbfit{n}})$ can be interpreted as a measure of the tendency of $\mathbfit{g}_p$ to align with a given direction.

The $\mathcal{D}$ statistic was originally applied to the four-year COBE DMR data, by choosing $\mathbfit{g}_p = \nabla T_p$ \citep{2000MNRAS.313..331B}. Since on large scales the \Planck\ temperature maps agree well with COBE
, we do not repeat this analysis. Instead, we first apply the statistic to polarization maps, with $\mathbfit{g}_p$ being the polarization field. Later, in Section~\ref{sec:lensing}, we apply the statistic to the \Planck\ lensing map, now letting $\mathbfit{g}_p$ be the field of lensing deflections.

For the polarization analysis, we express $\mathbfit{g}_p$ in terms of the Stokes parameters \textit{Q} and \textit{U}. First, the magnitude of $\mathbfit{g}_p$ is
\begin{equation}
   P =  \sqrt{Q^2 + U^2}.
   \label{P}
\end{equation}
The polarization direction is contained in the tangent plane to the celestial sphere at a given pixel \textit{p}. Following the CMB convention adopted by WMAP \citep{2007ApJS..170..335P} and \cite{2014A&A...571A...1P}, the angle of the polarization, $\gamma$, is measured from the meridian and taken to be positive for north through west. Then $\gamma$ is calculated as follows:
\begin{equation}
  \gamma =
  \begin{cases}
    \frac{1}{2}\arctan{\frac{U}{Q}}, & \text{if $Q \geq 0$ }; \\
    -\frac{\pi}{2}+\frac{1}{2}\arctan{\frac{U}{Q}}, & \text{if $Q < 0$ and $U < 0$};\\
    \frac{\pi}{2}+\frac{1}{2}\arctan{\frac{U}{Q}}, & \text{if $Q < 0$ and $U \geq 0$}.
  \end{cases}
  \label{gamma}
\end{equation}
Polarization is a spin-two quantity that is represented by headless pseudo-vectors; hence, $\gamma$ can be rotated by $180^\circ$ without changing the polarization. The quadratic definition of $f(\hat{\mathbfit{n}})$ allows us to treat the pseudo-vectors as regular vectors pointing in either direction.

At this point we could alternatively decompose polarization into the
(curl-free) $E$ and (divergence-free) $B$ modes
\citep[see e.g.][]{1997NewA....2..323H}. We could then choose to
examine directionality in the gradient of $E$, just as was done for the
gradient of $T$ in \citet{2000MNRAS.313..331B}; we could also do
the same thing for $B$ if it was non-zero.  We will not follow that path here.
However, we note in Appendix~\ref{sec:AppA} the slightly surprising result
that $\mathcal{D}$ can distinguish between $E$ modes and $B$ modes, and
Appendix~\ref{sec:AppB} further shows how $\mathcal{D}$ has sensitivity to
rotated polarization.

Returning to the use of $P$ and $\gamma$ to define the polarization field on
the sphere,
the weights $w_p$ must be chosen so that the noise structure and the masked sky do not introduce a preferred direction to $f(\hat{\mathbfit{n}})$. In other words, we want to choose the weights so that the ensemble-average $\langle f \rangle$ is constant as a function of $\hat{\mathbfit{n}}$ for a (possibly inhomogeneous) distribution of isotropic vectors $\mathbfit{g}_p$. We write equation~(\ref{f}) as
\begin{equation}
   f(\hat{\mathbfit{n}}) = \hat{\mathbfit{n}}^{\top} {\mathbfss A} \, \hat{\mathbfit{n}},
   \label{f'}
\end{equation}
where ${\mathbfss A}$ is the $3 \times 3$ matrix
\begin{equation}
   A_{ij} = \sum_{p=1}^{N} w_p g_{pi} g_{pj},
   \label{A}
\end{equation}
and $g_{pi}$ is the $i$th Cartesian coordinate of the vector $\mathbfit{g}_p$. Then requiring that $\langle f(\hat{\mathbfit{n}}) \rangle$ be independent of $\hat{\mathbfit{n}}$ is equivalent to requiring that $\langle{\mathbfss A}\rangle$ be proportional to the identity matrix. We have the freedom to normalize \textit{f} and we use that freedom to set ${\mathbfss A}$ equal to the identity, i.e.,
\begin{equation}
   \langle A_{ij} \rangle = \delta_{ij}.
   \label{A'}
\end{equation}
Since the ensemble average of ${\mathbfss A}$ can be written as
\begin{equation}
   \langle A_{ij} \rangle = \sum_{p=1}^{N} w_p \langle g_{pi} g_{pj} \rangle,
   \label{Aavg}
\end{equation}
equation~(\ref{A'}) constrains the weights $w_p$. To see this constraint in a more useful form, we use the assumption that $\mathbfit{g}_p$ is statistically isotropic. Let $\mathbfit{G}_p$ be a three-dimensional vector drawn from an isotropic distribution, and define $\mathbfit{g}_p$ to be the projection of $\mathbfit{G}_p$ onto the tangent plane of the sphere at pixel \textit{p}:
\begin{equation}
   \mathbfit{g}_p = \mathbfit{G}_p - (\mathbfit{G}_p \cdot \hat{\mathbfit{{r}}}_p) \, \hat{\mathbfit{r}}_p.
   \label{gp}
\end{equation}
This imposes the requirement that $\mathbfit{g}_p$ be isotropic in the tangent plane. Since $\mathbfit{G}_p$ is isotropic, $\langle {\mathbfit{G}_p} \rangle = 0$ and $\langle G_{pi} G_{pj} \rangle = P_p \delta_{ij}$, with $\mathit{P_p}$ being one third of the mean-squared amplitude of the vector $\mathbfit{G}_p$. Applying equation~(\ref{gp}) we obtain
\begin{align}
\begin{split}
   \langle g_{pi}g_{pj} \rangle = ~ & \langle G_{pi}G_{pj} \rangle - r_{pi}\sum_{\alpha=1}^{3} \langle G_{p\alpha}G_{pj} \rangle r_{p\alpha}\\
    &-r_{pj}\sum_{\beta=1}^{3} \langle G_{pi}G_{p\beta} \rangle r_{p\beta}\\
    & + \bigg( \sum_{\alpha,\beta=1}^{3} \langle G_{p\alpha}G_{p\beta} \rangle r_{p\alpha}r_{p\beta} \bigg) r_{pi}r_{pj}\\
    = ~ & P_p(\delta_{ij}-r_{pi}r_{pj}).
    \label{gpavg}
\end{split}
\end{align}
Combining equation~(\ref{gpavg}) with equations~(\ref{Aavg}) and~(\ref{A'}), we obtain
\begin{equation}
   \delta_{ij} = \sum_{p=1}^{N} w_p P_p Q_{pij},
   \label{delta}
\end{equation}
where
\begin{equation}
   Q_{pij} = \delta_{ij}-r_{pi}r_{pj}.
   \label{Qpij}
\end{equation}
Since equation~(\ref{delta}) is symmetric, we have six constraints on the $N$ pixel weights $w_p$. The choice of  weights is therefore very underdetermined, and we need additional criteria to specify them. One natural criterion is that the weights should be as nearly equal as possible. That would mean minimizing the variance of {$w_p$}. However, it is easier to minimize the variance of $\tilde{w}_p \equiv w_pP_p$, so we do that instead. Therefore, we would like to minimize
\begin{equation}
   \mathrm{Var}(\tilde{w}_p) = \frac{1}{N} \sum_{p=1}^{N} \tilde{w}_p^2 - {\bigg( \frac{1}{N} \sum_{p=1}^{N} \tilde{w}_p \bigg)}^2.
   \label{Varw}
\end{equation}
Taking the trace of equation~(\ref{delta}), we see that the second term in equation~(\ref{Varw}) is constant, since $\sum_{p=1}^{N} \tilde{w}_p = \frac{3}{2}$. Hence,
\begin{equation}
   \Delta^2 \equiv \frac{1}{2} \sum_{p=1}^{N} \tilde{w}_p^2
   \label{Delta}
\end{equation}
must be minimized subject to the constraint of equation~(\ref{delta}). Introducing $\Lambda$, a symmetric $3 \times 3$ matrix of Lagrange multipliers, the problem may be written as
\begin{equation}
  \tilde{w}_p = \sum_{i,j=1}^{3} \Lambda_{ij} Q_{pij}.
  \label{w'}
\end{equation}
Substituting equation~(\ref{w'}) back into equation~(\ref{delta}), we obtain
\begin{equation}
  \delta_{ij} = \sum_{k,l=1}^{3} \Lambda_{kl} \tilde{Q}_{ijkl},
  \label{delta'}
\end{equation}
with
\begin{equation}
  \tilde{Q}_{ijkl} = \sum_{p=1}^{N} Q_{pij} Q_{pkl}.
  \label{Qijkl}
\end{equation}
This is a six-dimensional linear system, solvable for $\Lambda$. After finding $\Lambda$, the weights $\tilde{w}_p$ are easily calculated using equation~(\ref{w'}).

Now that we have $\tilde{w}_p$ we can calculate $w_p$ using the definition $\tilde{w}_p \equiv w_pP_p$.
However, we must first calculate $P_p$, which is the mean-squared amplitude of a Cartesian component of the vector $\mathbfit{G}_p$. Since $\mathbfit{g}_p$ is the projection of the isotropic vector $\mathbfit{G}_p$ onto the tangent plane of the sphere, we can express $P_p$ as
\begin{equation}
   P_p = \frac{1}{2} \langle \mathbfit{g}_p \cdot \mathbfit{g}_p  \rangle.
   \label{Pp}
\end{equation}
Hence, the value of $P_p$ is proportional to the mean square amplitude of the polarization pseudo-vectors at pixel $p$ for the simulations of the map being investigated. For the case of polarization maps, variations of $P_p$ from pixel to pixel are due to the noise structure of the observations, since the assumed signal variance is the same at all pixels.

After calculating the weights, finding the $\mathcal{D}$ statistic is computationally very quick. The maximum and minimum values of $f(\hat{\mathbfit {n}})$ subject to the constraint $\sum_{i=1}^{3} \hat{n}^2_i = 1$ can be solved by introducing a Lagrange multiplier $\lambda$. For the Cartesian components of $\hat {\mathbfit {n}}$, we set the derivative of \textit{f} with respect to $\hat{n}_i$ equal to the derivative of the constraint equation multiplied by $\lambda$. This gives us
\begin{equation}
  2 \sum_{j=1}^{3} A_{ij} \hat{n}_i = 2 \lambda\hat{n}_i,
  \label{Eigen1}
\end{equation}
which can be written in matrix form as
\begin{equation}
  \mathbfss A \hat{\mathbfit{n}} = \lambda  \hat{\mathbfit{n}}.
  \label{Eigen2}
\end{equation}
Now we see that the locations of the extrema of \textit{f} are the eigenvectors of $\mathbfss{A}$ and the extreme values are given by the eigenvalues of $\mathbfss{A}$. Since $\mathbfss A$ is symmetric, it must have three real eigenvectors, and so \textit{f} has three critical points, which are a maximum, a minimum and a saddle. After computing the elements of $\mathbfss A$, $\mathcal{D}$ can be calculated as the largest eigenvalue of $\mathbfss A$ divided by the smallest eigenvalue. The maximal and minimal directions of the map are given by the eigenvectors corresponding to the largest and smallest eigenvalues, respectively.

Once $\mathcal{D}$ has been calculated for real sky data, we can compare its value to that found for simulations of the CMB and noise. Calculating $\mathcal{D}$ for a large number of these simulations gives a distribution of values, and excess directionality in a CMB data set appears as a value of $\mathcal{D}$ that is an outlier of the distribution.

$\mathcal{D}$ is a very simple statistic for identification of statistical anisotropy in a CMB map. Since it can be calculated in $O(N)$ operations, its speed makes it appropriate to include in any tool-kit for looking at the statistical isotropy of CMB maps.

\section{Results}
\label{sec:results}

In this section, we  describe the results of applying the $\mathcal{D}$ statistic to \textit{Planck} polarization and lensing maps. Specifically, in the following six subsections, we will show results for polarized synchrotron and dust foregrounds, single-frequency maps, CMB maps and lensing deflection. The general procedure for analysing a map's directionality using the $\mathcal{D}$ statistic is:
\begin{enumerate}
\item obtain the map to be analysed and define $\mathbf{\mathit{g_p}}$ with respect to its data type;
\item find (or create) an appropriate mask;
\item create simulations of the map;
\item using the mask and the simulations, calculate appropriate weights, $w_p$, as described in Section~\ref{sec:dstat};
\item using the weights, calculate the $\mathcal{D}$ statistic for the simulations, as well as for the original map;
\item compare the value of $\mathcal{D}$ calculated for the original map to the distribution calculated for the simulations.
\end{enumerate}
Relevant details of this procedure will be discussed in each subsection.

Since we are only interested in relatively large-angle behaviour, it will be convenient to degrade the resolution of the maps. We choose {\tt HEALPix} $N_{\rm side}=16$ \citep[see][]{2005ApJ...622..759G}. This resolution is sufficient to encompass the large-scale polarization pattern and allows us to quickly simulate maps and calculate the $\mathcal{D}$ statistic for those simulations.

It is worth remembering that the reason we can use the directionality of a CMB map as a proxy for Galactic foregrounds lies in the fact that these foregrounds introduce directionality to the intensity and polarization along the axis of the Galactic poles. This effect on directionality will be demonstrated in Section~\ref{sec:foregrounds}; but first, as an example of analysing maps for directionality using the $\mathcal{D}$ statistic we investigate the \Planck\ polarization maps.

\subsection{Analysis of \textit{Planck} \textit{2018} frequency maps}
\label{sec:freq}

The latest maps from the Planck Collaboration are from the 2018 release
(``PR3''), with basic data reduction procedures described in
\citet{2018planckII} and \citet{2018planckIII}.
We use the \textit{Q} and \textit{U} polarization maps to define the local directionality vector $\mathbf{\mathit{g_p}}$, as shown in equations~(\ref{P}) and (\ref{gamma}). We first calculate the $\mathcal{D}$ statistic for the single-frequency maps, comparing its value to the distribution calculated for noise and CMB simulations.

Degrading the maps makes the analysis faster, and we do this by using {\tt HEALPix} routine \texttt{alm2map}. As done in \citet{collaboration2015planck}, we apply a Gaussian beam with a full width at half maximum (FWHM) specified by the degraded resolution of the map. For the degraded resolution of $N_{\rm side} = 16$, we use FWHM of 160 arcmin. The next step before calculating the $\mathcal{D}$ statistic is to mask the sky map with the GAL070 mask \citep{2016A&A...594A...8P}. Before application, the mask must also be degraded to the same resolution; to do this we use the \texttt{ud\_grade} function in \texttt{HEALPix} and assign the value 0 to all pixels with values less then 0.9 in the degraded map, with all other pixels given the value 1.

To create simulations for each frequency, we use the theoretical angular power spectrum for the best-fit $\Lambda$CDM model provided in \cite{2016A&A...594A..13P}, as well as the covariance matrices provided with each frequency map. From {\tt HEALPix}, \texttt{synfast} was used to make the CMB signal simulations at the degraded resolution. The covariance matrices are provided with a resolution corresponding to $N_{\rm side}=2048$ for High Frequency Instrument maps and $N_{\rm side}=1024$ for Low Frequency Instrument maps. We generate correlated, inhomogeneous noise simulations at the same resolutions by using the Cholesky decomposition of the covariance matrices. After generating a noise simulation, it is also degraded to the same resolution and combined with the CMB signal. Finally, the simulation is masked as described above, for consistency with the actual data that it will be compared to.

\begin{figure}
\begin{center}\vbox{
  \includegraphics[width=\columnwidth]{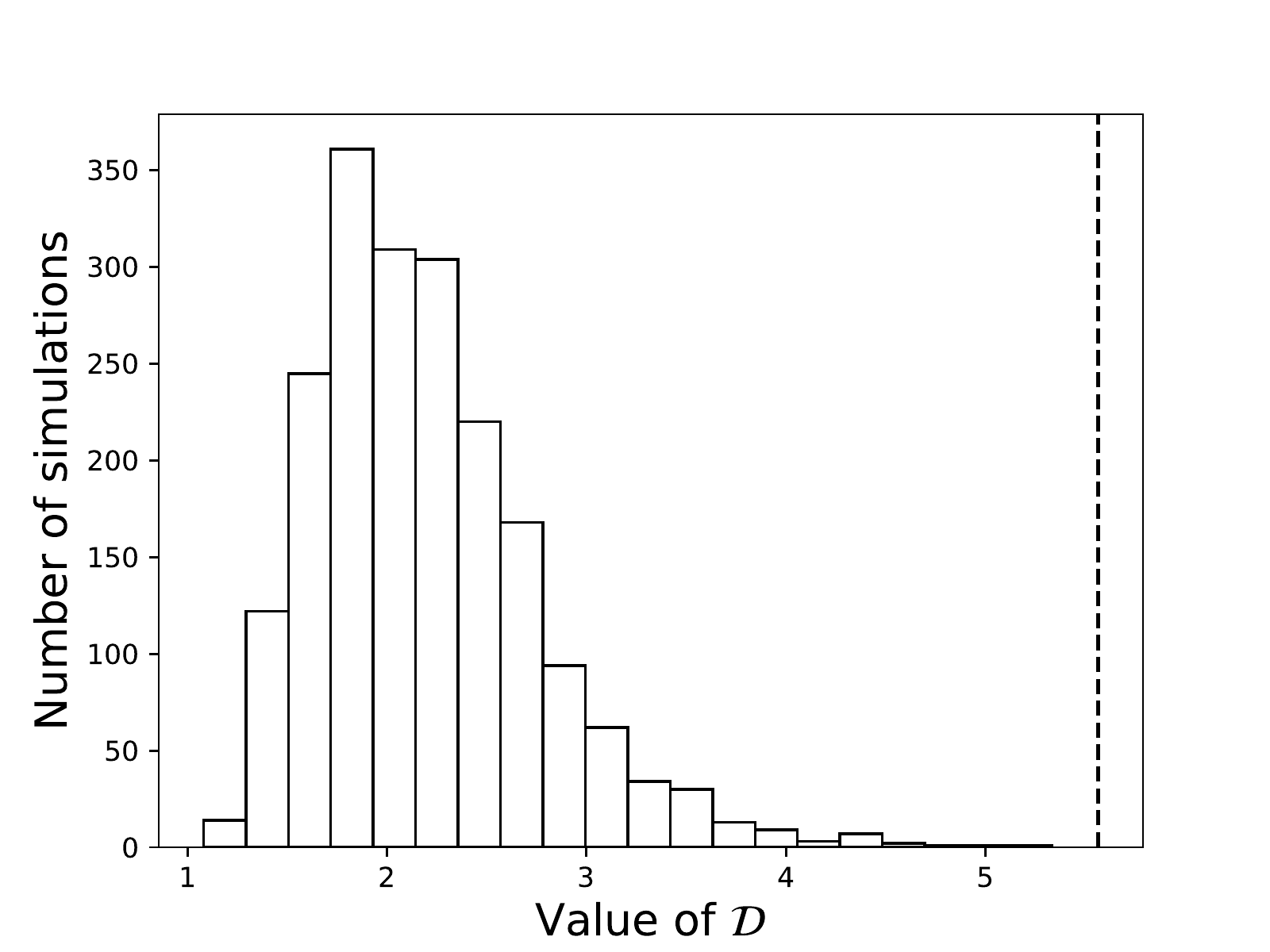}}
\end{center}
    \caption{Directionality histogram for simulated 70-GHz maps. The dashed line, shown at $\mathcal{D} = 1.12$, is the value calculated for the actual 70-GHz data. All maps and simulations shown here have been degraded to $N_{\rm side}=16$ and masked with the GAL070 mask.
    \label{fig:1}}
\end{figure}

For each frequency we generate and analyse 2000 simulations and compare the distribution of their $\mathcal{D}$ statistics to that calculated for the actual data (degraded and masked as described above). An example of this is shown in Fig.~\ref{fig:1}.

\begin{figure}
  \includegraphics[width=\columnwidth]{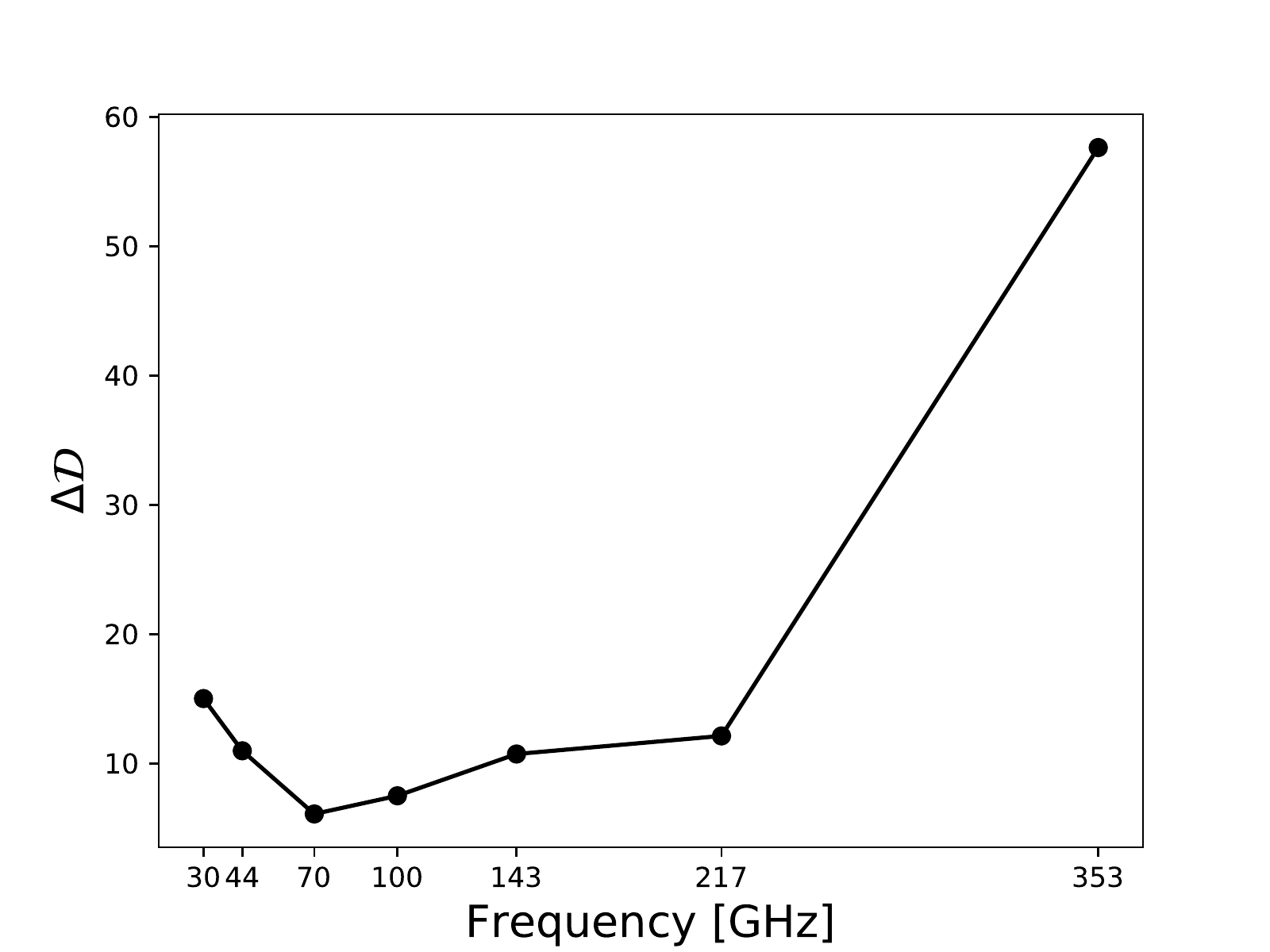}
    \caption{Distance of the $\mathcal{D}$ statistic (in numbers of $\sigma$) for each frequency map from the mean for their respective foreground-free simulations. The maps and simulations have been masked with the GAL070 mask and degraded to $N_{\rm side}=16$.}
    \label{fig:nsides}
\end{figure}

We use the distance from the mean (in units of $\sigma$) as a measure of the significance with which we detect foregrounds. Specifically,
\begin{equation}
  \Delta \mathcal{D} \equiv \frac{|\mathcal{D} - \bar{\mathcal{D}}|}{\sigma},
  \label{DistMean}
\end{equation}
where $\mathcal{D}$ is the value calculated for the real sky data, $\bar{\mathcal{D}}$ is the mean value of $\mathcal{D}$ calculated for simulations and $\sigma$ is the standard deviation of $\mathcal{D}$ for the simulations. Figure~\ref{fig:nsides} shows $\Delta D$ calculated for all frequencies at $N_{\rm side}=16$.

\begin{figure*}
  \begin{tabular}{@{}cc@{}}
    \includegraphics[width=0.85\columnwidth]{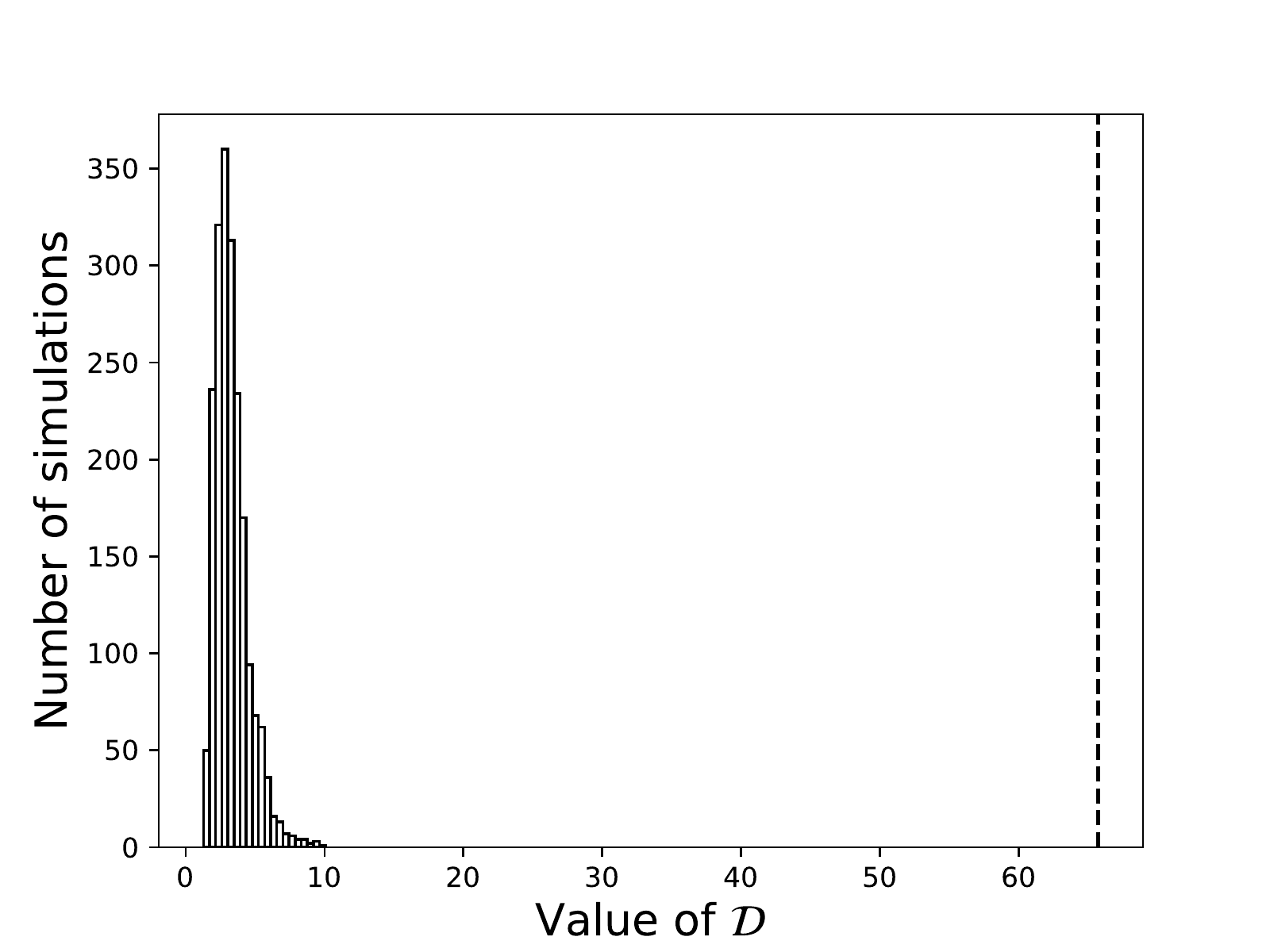} &
    \includegraphics[width=0.85\columnwidth]{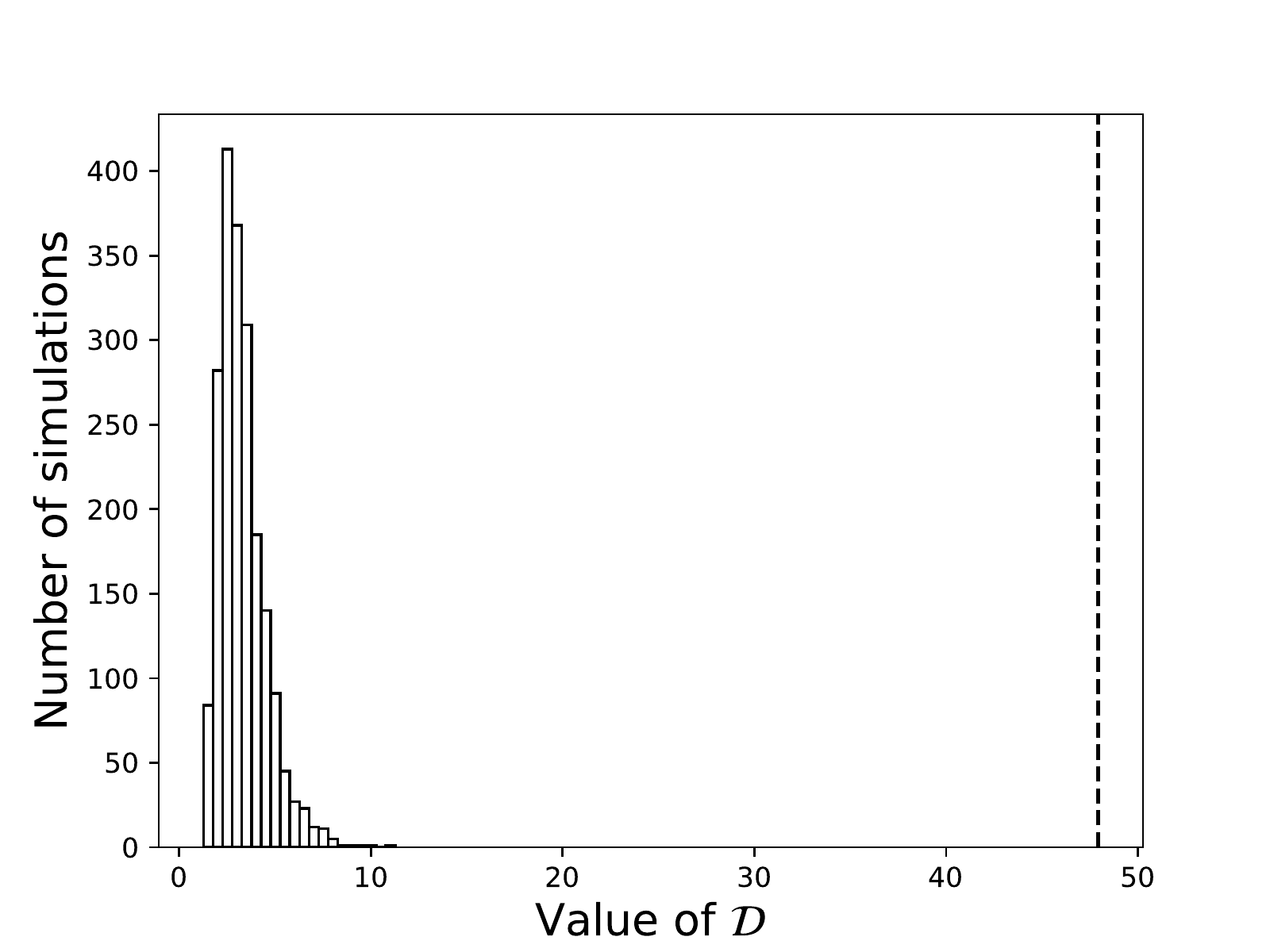} \\
    \includegraphics[width=0.85\columnwidth]{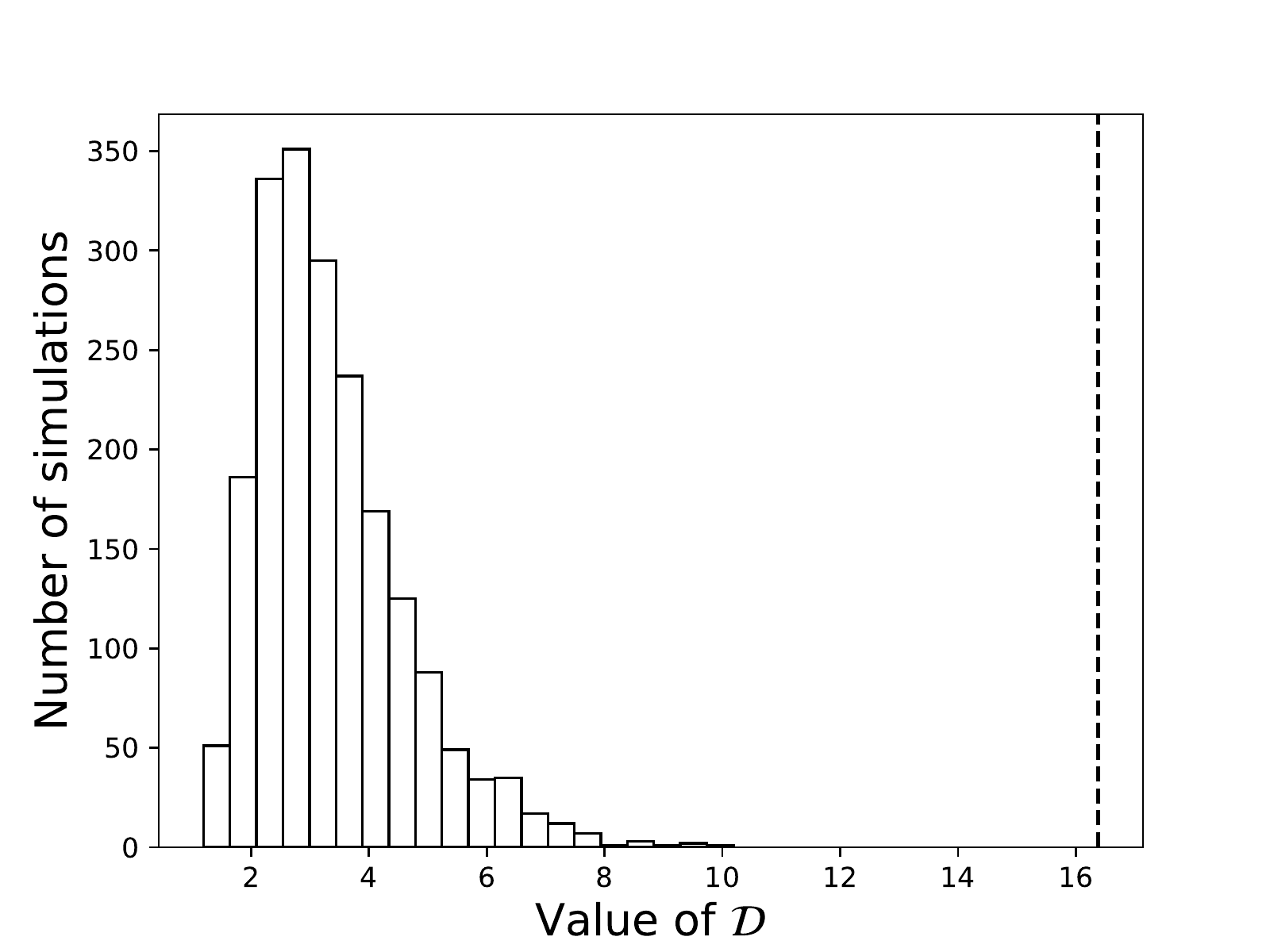} &
    \includegraphics[width=0.85\columnwidth]{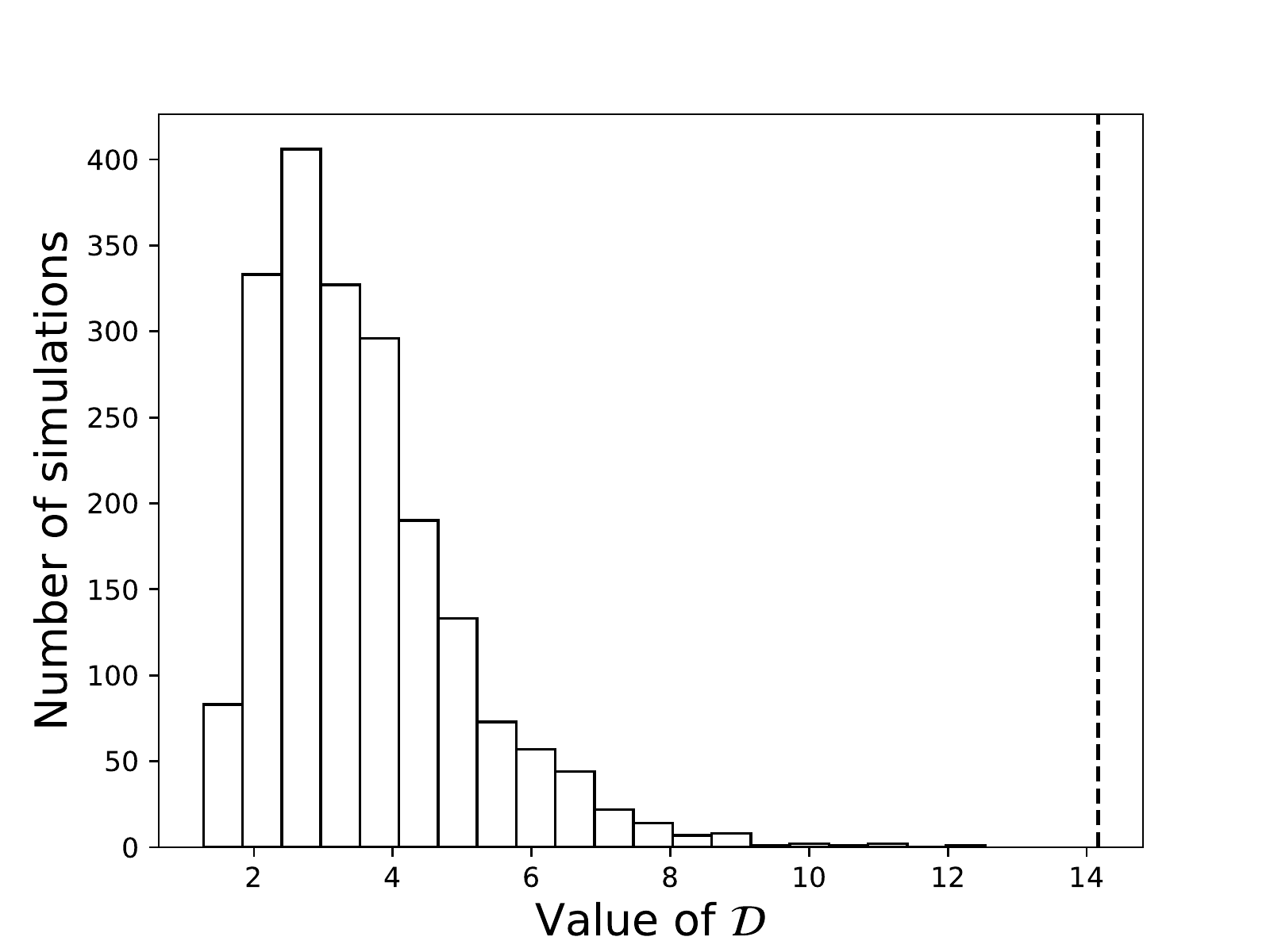} \\
    \includegraphics[width=0.85\columnwidth]{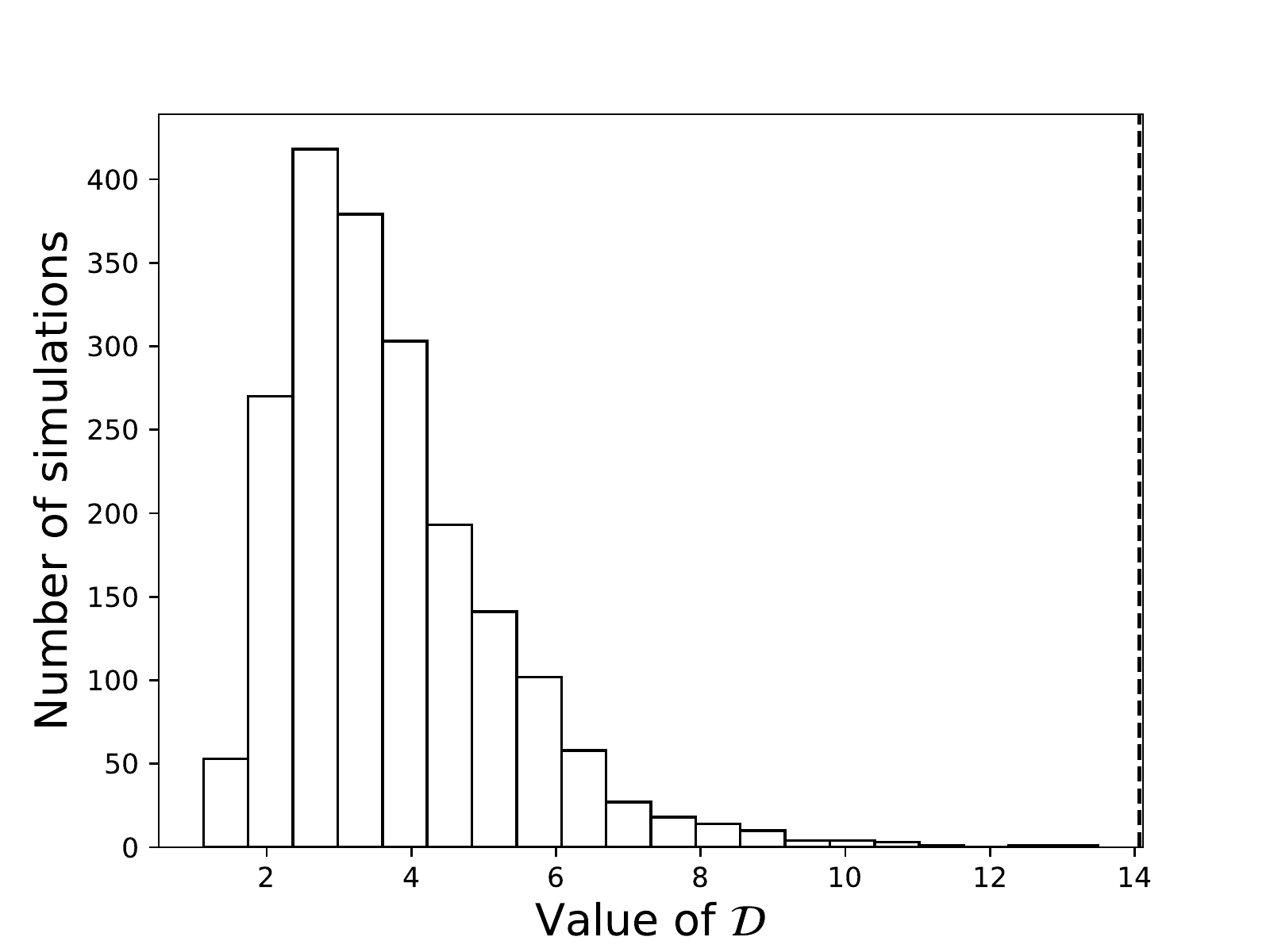} &
    \includegraphics[width=0.85\columnwidth]{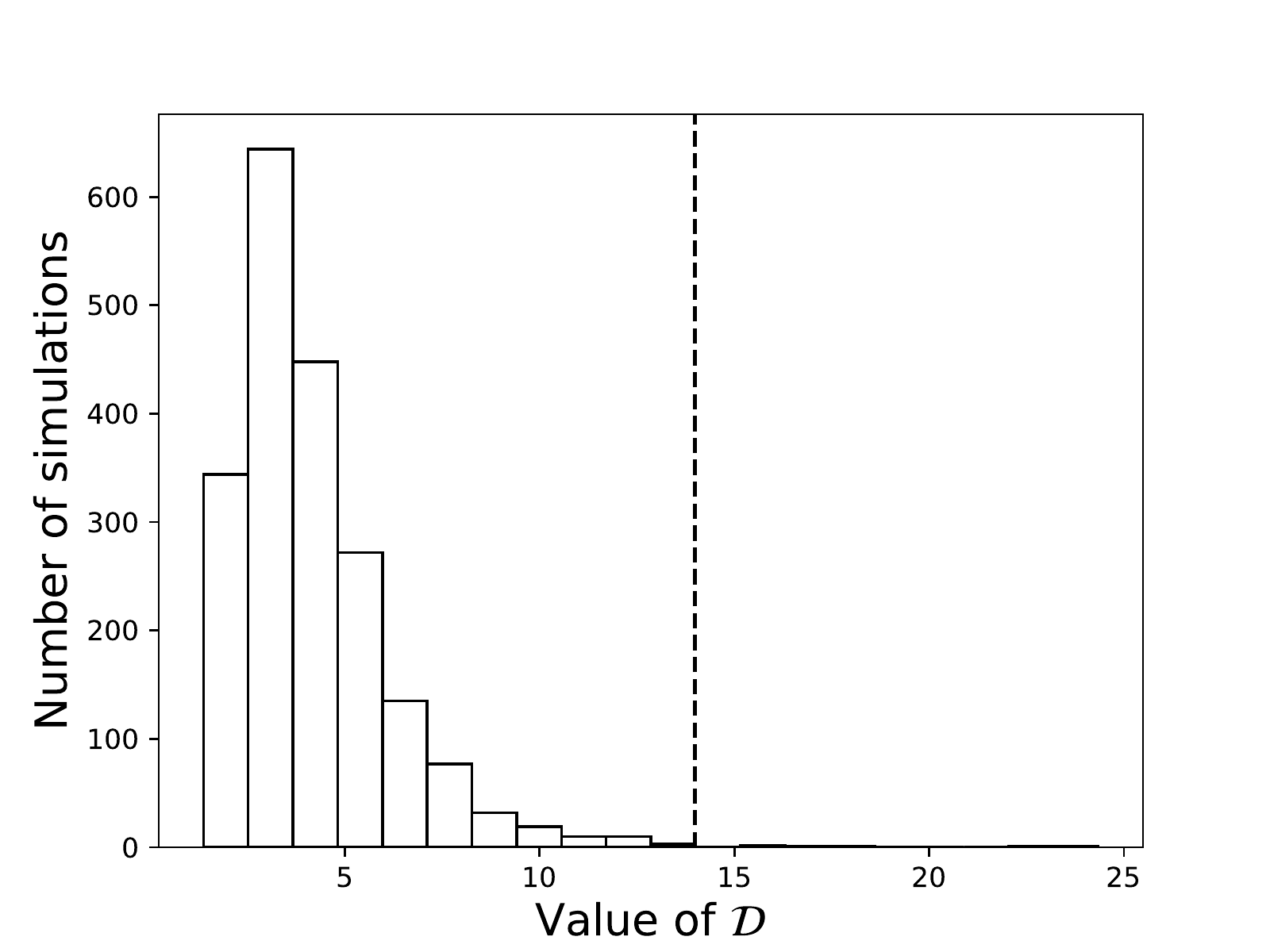} \\
    \includegraphics[width=0.85\columnwidth]{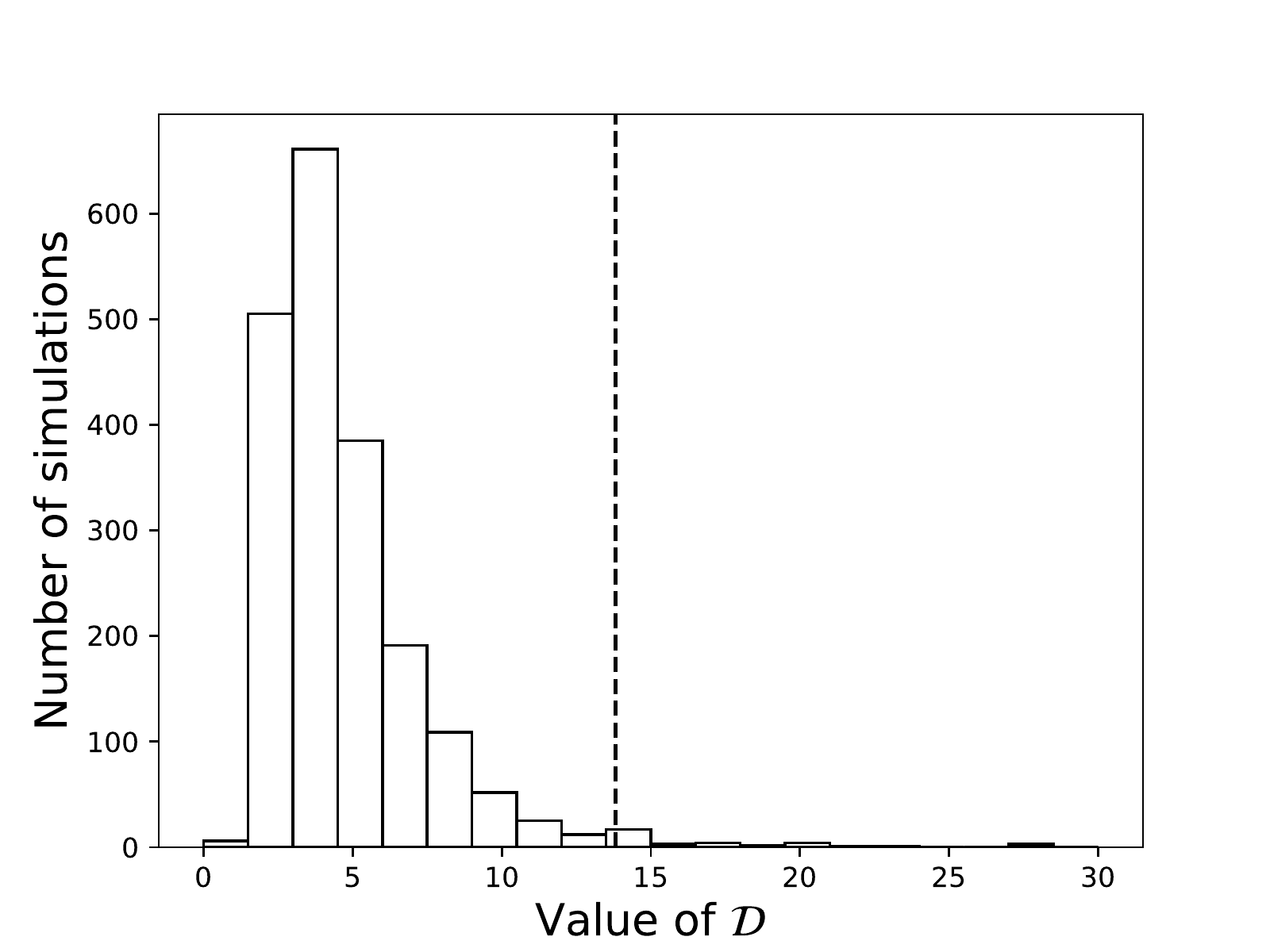} &
    \includegraphics[width=0.85\columnwidth]{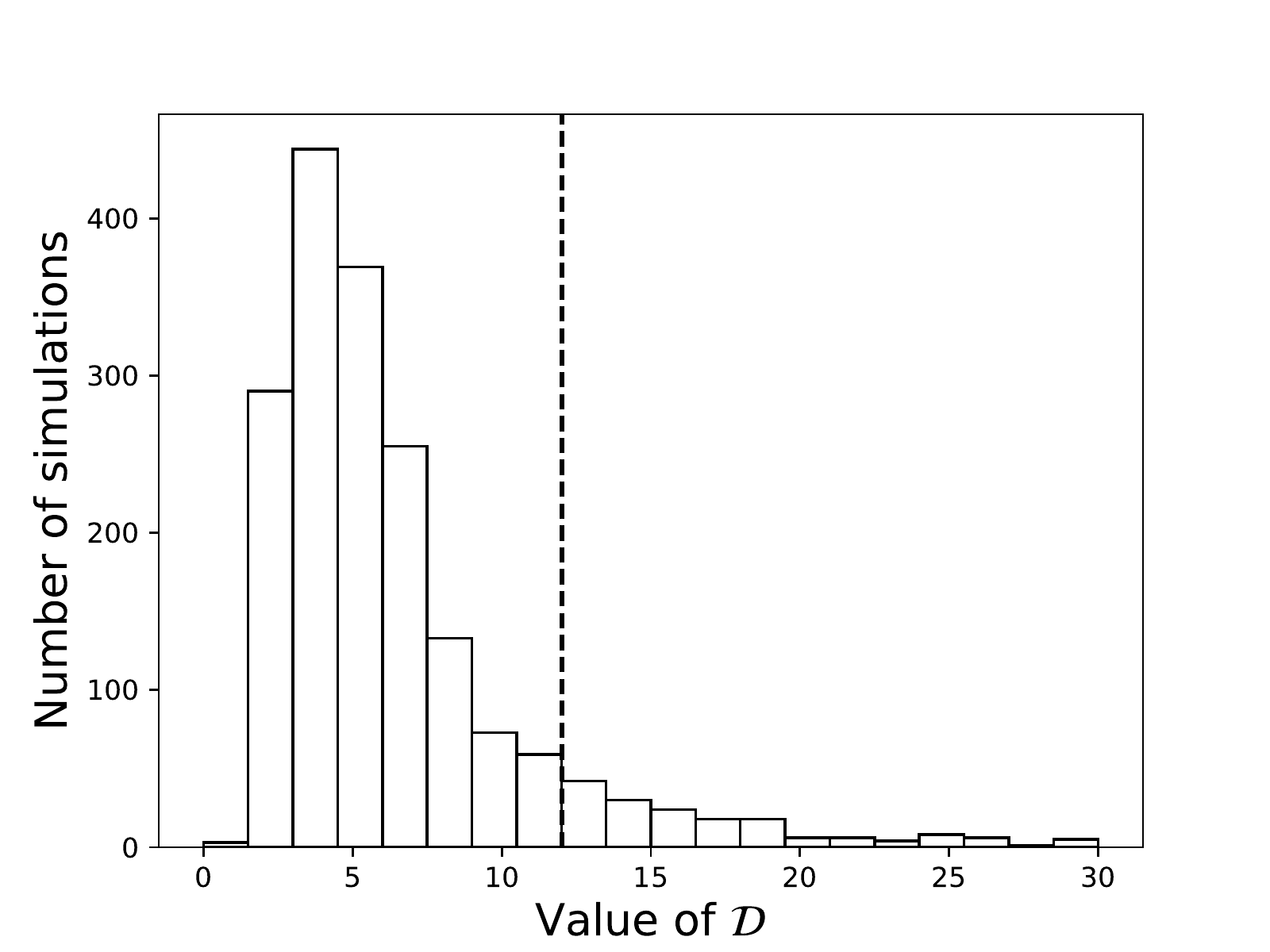} \\
  \end{tabular}
  \caption{$\mathcal{D}$ statistic for the 100-GHz data (dashed line) and the distribution of $\mathcal{D}$ for simulations, starting with no mask and repeated for a series of increasingly large masks. As the width of the mask is increased the $\mathcal{D}$ statistic falls into the distribution of the simulations. All maps and simulations shown here have been degraded to $N_{\rm side}=16$. The mask thickness (in degrees measured from the Galactic plane) for each figure is $0^\circ$, $8^\circ$, $24^\circ$, $40^\circ$, $48^\circ$, $56^\circ$, $64^\circ$ and $72^\circ$ (reading from left to right and top to bottom).}
  \label{fig:3}
\end{figure*}

\begin{figure}
  \includegraphics[width=\columnwidth]{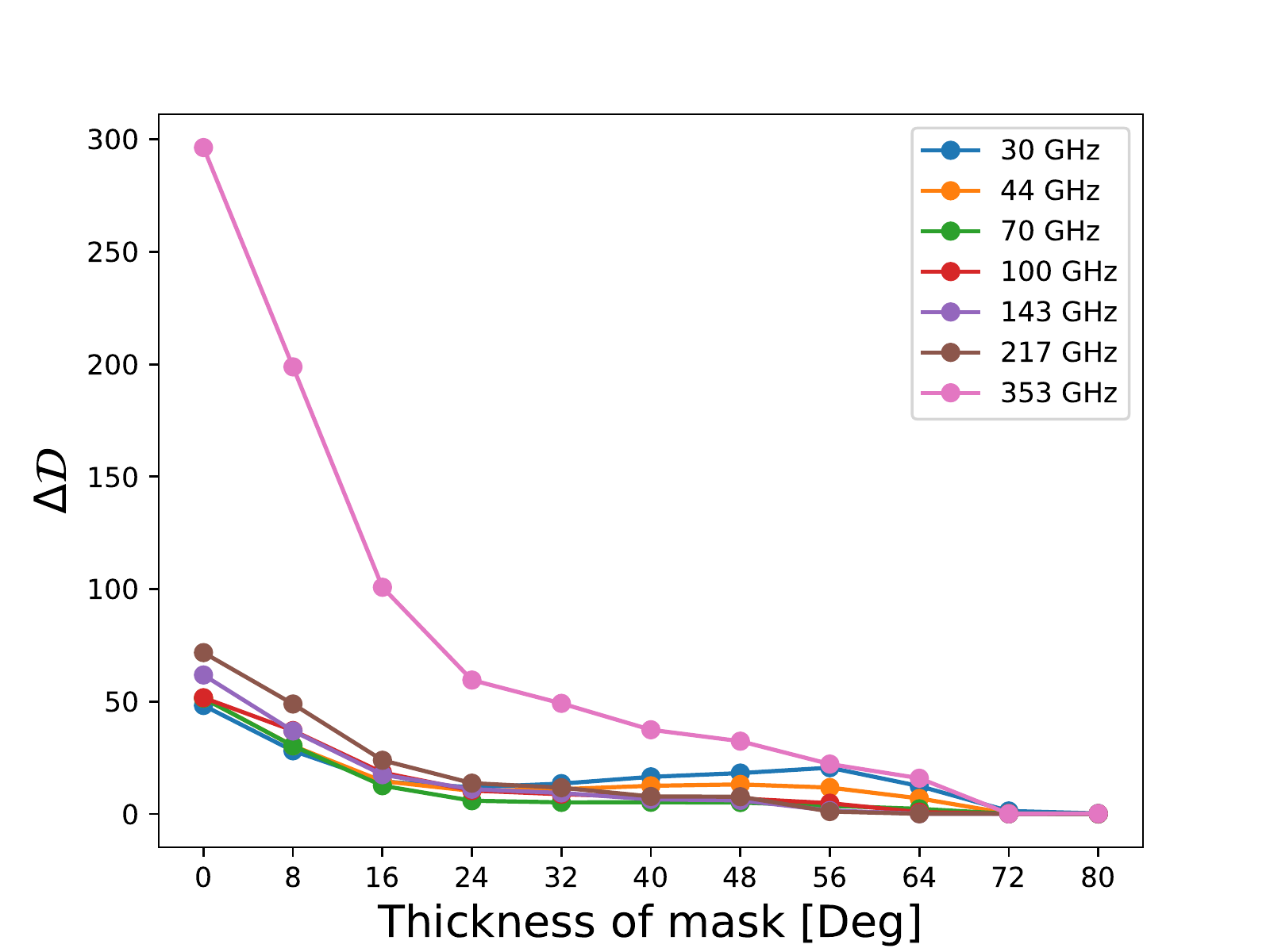} \\
  \caption{Distance of the $\mathcal{D}$ statistic, for each frequency map, from the mean for their respective foreground-free simulations, displayed as a function of mask thickness for all frequency maps and measured in numbers of $\sigma$. Figure~\ref{fig:3} displays the value of the $\mathcal{D}$ statistic with respect to the distributions of foreground free-simulations for each mask thickness for the particular case of the 100-GHz frequency channel. The maps and simulations have been degraded to $N_{\rm side}=16$.}
  \label{fig:4}
\end{figure}

Returning to our tests on the frequency maps, we repeat this procedure of calculating the $\mathcal{D}$ statistic for actual data and comparing it to the distribution for simulated data, only this time we vary the mask. We start with no mask and for each iteration we increment the thickness of the mask. Using \texttt{pix2ang}, we mask all pixels that are within $0^\circ,4^\circ,8^\circ,12^\circ,16^\circ,20^\circ,24^\circ,28^\circ,32^\circ,36^\circ$ and $40^\circ$ of zero Galactic latitude. An example of this for the 100-GHz map is shown in Fig.~\ref{fig:3}. As the thickness of the mask is increased, more of the Galactic plane is cut out and so we expect less foreground contamination, which results in less directionality. This is indeed what we see in Fig.~\ref{fig:4}, which displays $\Delta \mathcal{D}$ as a function of the thickness of the mask for all frequencies.

\subsection{Analysis of \textit{Planck} \textit{2018} CMB maps}
\label{sec:cmb}

We now move our focus to full CMB maps. In this case, we use the \texttt{dx12\_v3} Monte Carlo simulations provided by the Planck Legacy Archive (PLA\footnote{\url{http://pla.esac.esa.int}}). There are 300 simulations provided for each of the four component-separation codes. As done in Section~\ref{sec:freq}, we degrade all of these simulations to $N_{\rm side}=16$. Before calculating the $\mathcal{D}$ statistic, maps are masked with the UP78 mask described in \citet{2016A&A...594A...9P}. The UP78 mask is degraded with the same method used to degrade the GAL070 mask.

\begin{figure}
  \includegraphics[width=\columnwidth]{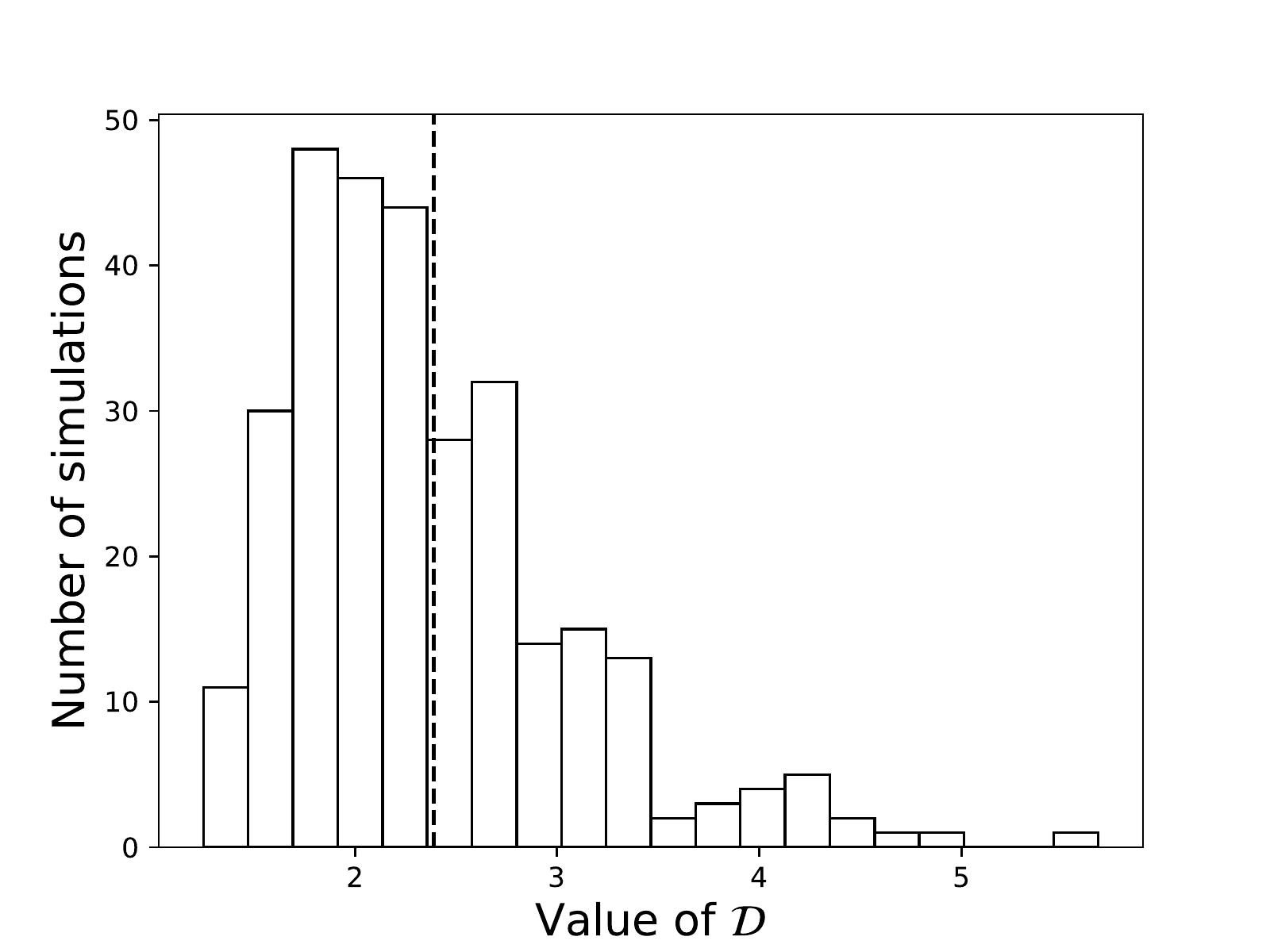} \\
  \caption{Directionality histogram for simulations of the \texttt{Commander} map. The dashed lines indicate the value of $\mathcal{D}$ calculated for the data. Both simulations and the real map are masked with the UP78 mask provided by the Planck Collaboration. All maps are degraded to $N_{\rm side}=16$.}
  \label{fig:5}
\end{figure}

Using the \texttt{dx12\_v3} Monte Carlo simulations we can analyse full CMB maps for directionality. In Table~\ref{tab:2} the $\Delta \mathcal{D}$ values and the \textit{p}-values are listed for each map. The bottom plot in Fig.~\ref{fig:5} shows the directionality distribution and $\mathcal{D}$ value for the \texttt{Commander} map.

\begin{table}
  \centering
  \caption{Distance of the $\mathcal{D}$ statistic for each CMB map from the mean for their respective foreground-free simulations. The  \textit{p}-value is also given for each map. The maps and simulations have been masked with the UP78 mask and degraded to $N_{\rm side}=16$.}
  \label{tab:2}
  \begin{tabular}{cccc}
    \hline
    Map & $\Delta \mathcal{D}$ & \textit{p}-value\\
    \hline
    \texttt{SMICA} & 1.917 &  0.053\\
    \texttt{Commander} & 0.045 & 0.386\\
    \texttt{NILC} & 1.779 & 0.06\\
    \texttt{SEVEM} & 1.300 & 0.103\\
    \hline
  \end{tabular}
\end{table}

In Table~\ref{tab:2}, we see no significant evidence of foreground contamination in any of the four maps. We do note that, with respect to its simulations, the \texttt{Commander} map shows the least signs of directionality.

\subsection{Directionality of polarization foregrounds}
\label{sec:foregrounds}

Galactic synchrotron and thermal dust emission are the two main sources of contamination in CMB polarization maps. As cosmic-ray electrons orbit in the Galactic magnetic field their acceleration causes them to emit synchrotron radiation, polarized preferentially towards the Galactic north \citep{2016planckXXV}. Polarized dust emission results from non-spherical dust grains that tend to align their long axes perpendicular to the magnetic field and preferentially emit radiation polarized along their long axes \citep{1951ApJ...114..206D}; this also gives large-scale directionality in the Galaxy \citep{2018planckXII}.

To demonstrate the directionality of these two sources of contamination, we analyse the polarized thermal dust emission and polarized synchrotron emission foreground maps described in \citet{2016A&A...594A..10P} and \citet{2018planckIV}. The general procedure for analysing a map for directionality using the $\mathcal{D}$ statistic involves creating appropriate simulations; for this particular goal of demonstrating that foregrounds give directionality towards the Galactic poles it is sufficient to use simulations of pure CMB skies plus noise.
In all methods, maps are degraded to $N_{\rm side}=16$, as described in Section~\ref{sec:freq}. The results are summarized in Table~\ref{tab:angles}.

For the first method we simply ignore the effects of the noise structure and use a uniform weighting scheme. Doing so we find that the maximal directions are $1.67^\circ$ and $0.71^\circ$ away from the Galactic poles for the polarized synchrotron and dust maps, respectively.

This first method has a potential flaw, namely that it ignores the fact that the noise is inhomogeneous, with lower noise near the Ecliptic poles. This inhomogeneity could introduce a false positive detection of directionality. In Section~\ref{sec:cmb}, we described a method for producing noise simulations that mimic the inhomogeneous noise structure. To assess whether this matters for the foreground maps, we now adopt a second analysis method in which we use the inhomogeneous $P_p$ values obtained for the \texttt{Commander} map in Section~\ref{sec:cmb}, to determine the weights. With this method, the angles between the maximal directions and Galactic poles are $2.81^\circ$ and $0.62^\circ$ for the polarized synchrotron and dust maps, respectively.

\begin{table}
  \centering
  \caption{Angle between the Galactic poles and the maximal direction obtained for the Galactic synchrotron and Galactic dust emission polarization maps. Two different methods are used to obtain the maximal direction, both described in Section~\ref{sec:foregrounds}.
}
  \label{tab:angles}
  \begin{tabular}{ccccc}
    \hline
    Method no. & Angle for synchrotron & Angle for dust\\
    \hline
    1 &  $2^\circ$  &  $1^\circ$ \\
    2 & $3^\circ$ & $1^\circ$\\
    \hline
  \end{tabular}
\end{table}

\begin{figure}
  \includegraphics[width=\columnwidth]{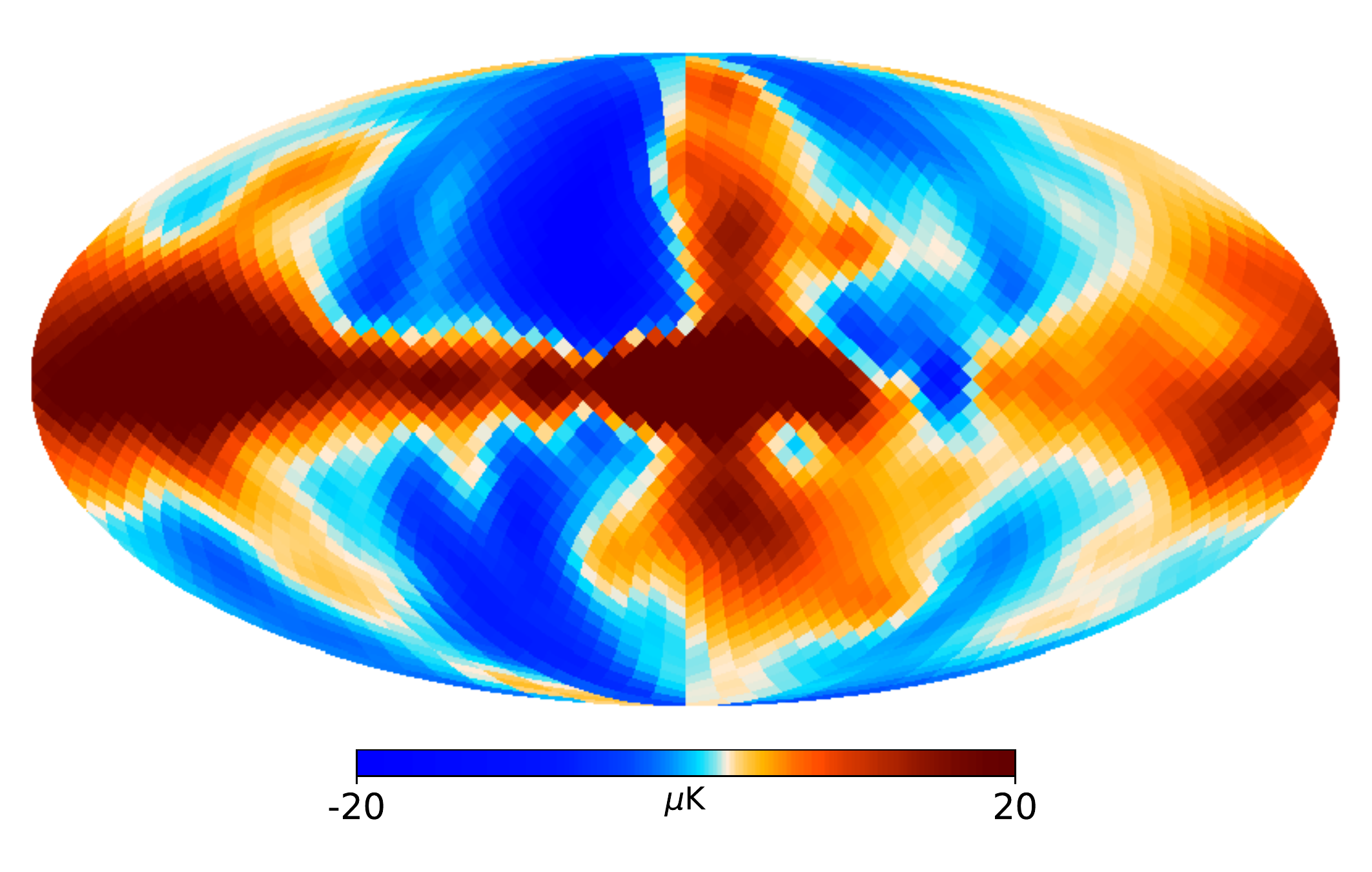}
  \includegraphics[width=\columnwidth]{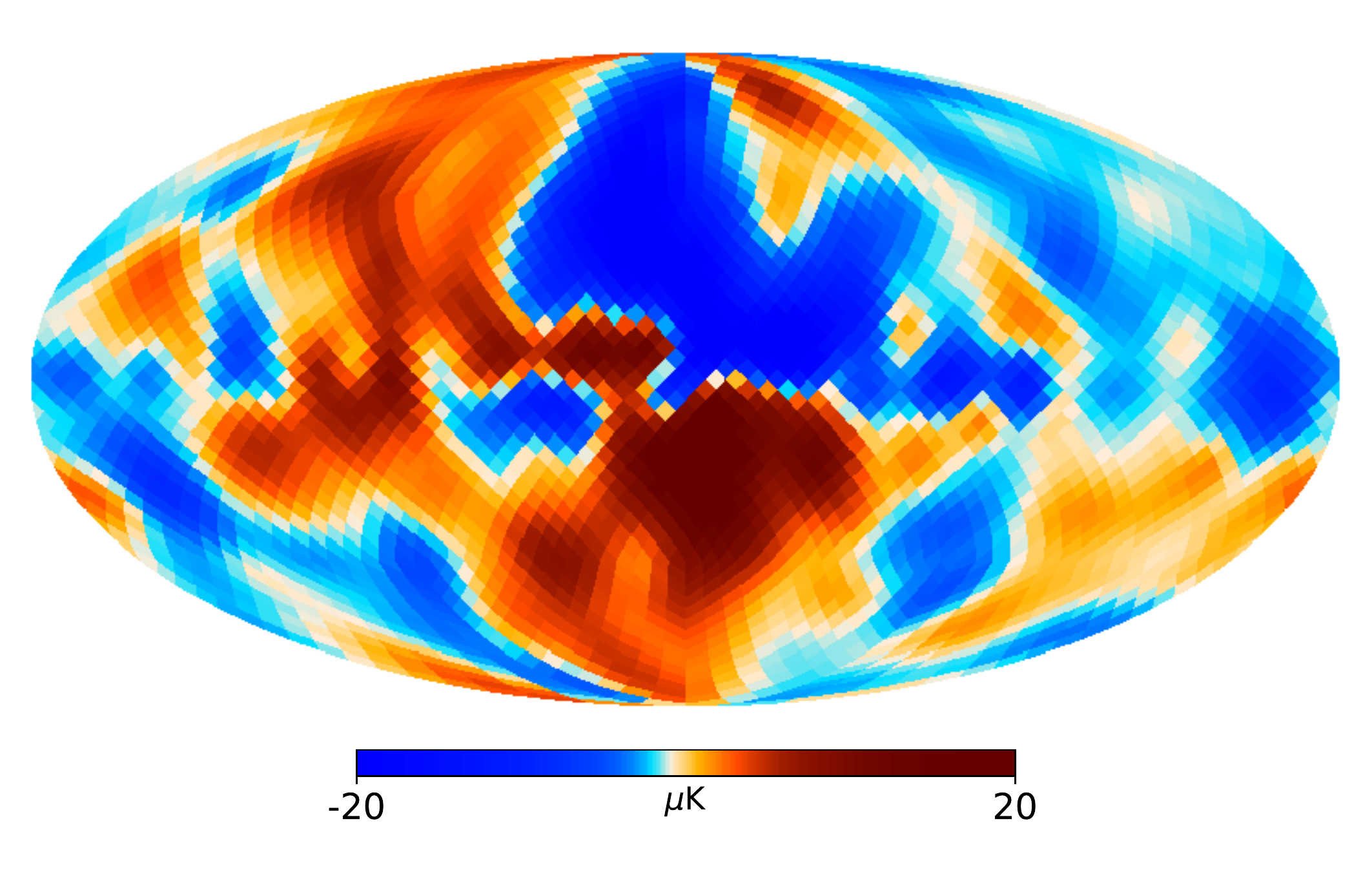}
    \caption{Stokes $Q$ (top) and $U$ (bottom) maps for the synchrotron foreground component at resolution $N_{\rm side}=16$.}
    \label{SQU1}
\end{figure}

\begin{figure}
  \includegraphics[width=\columnwidth]{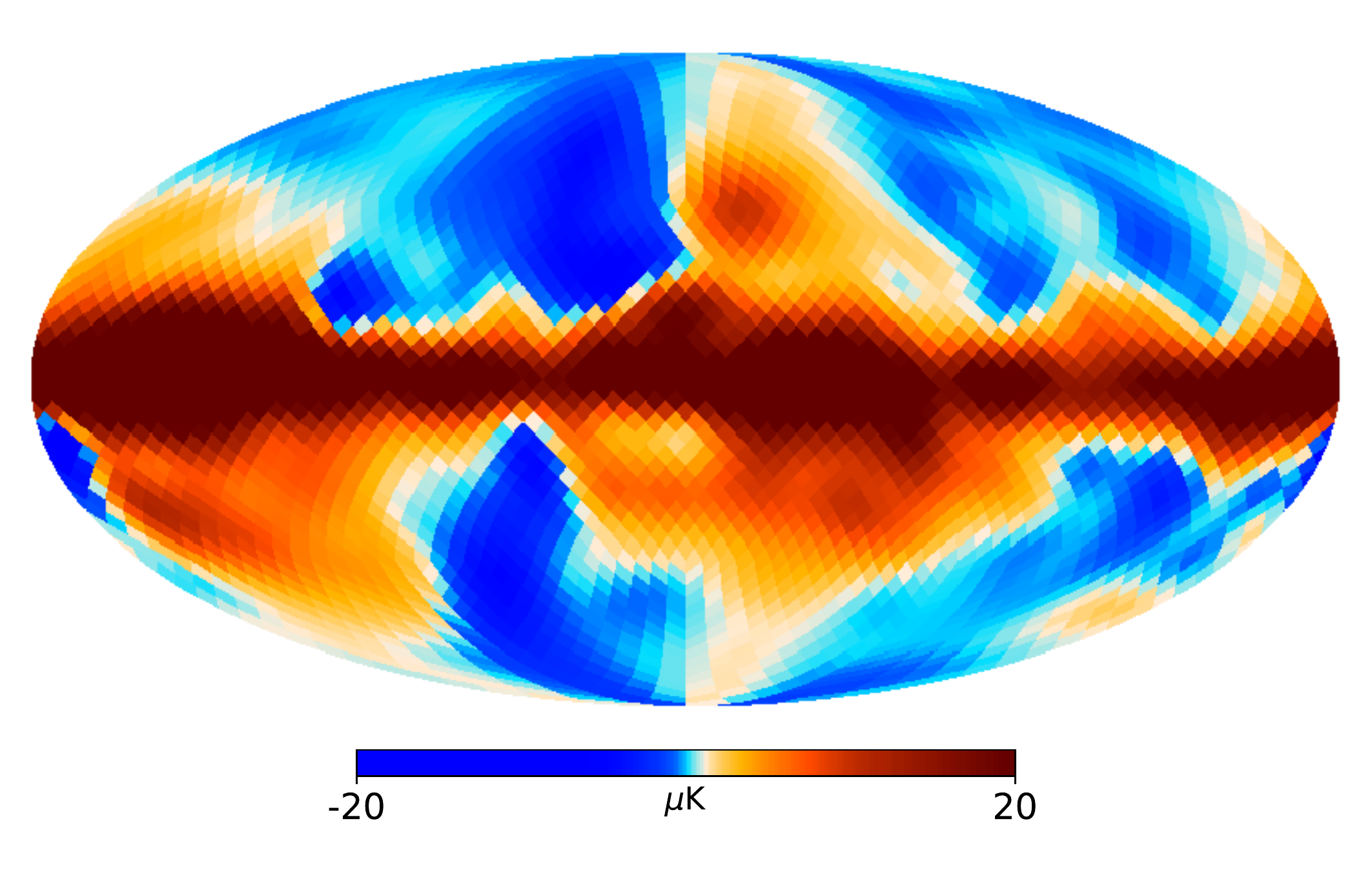}
  \includegraphics[width=\columnwidth]{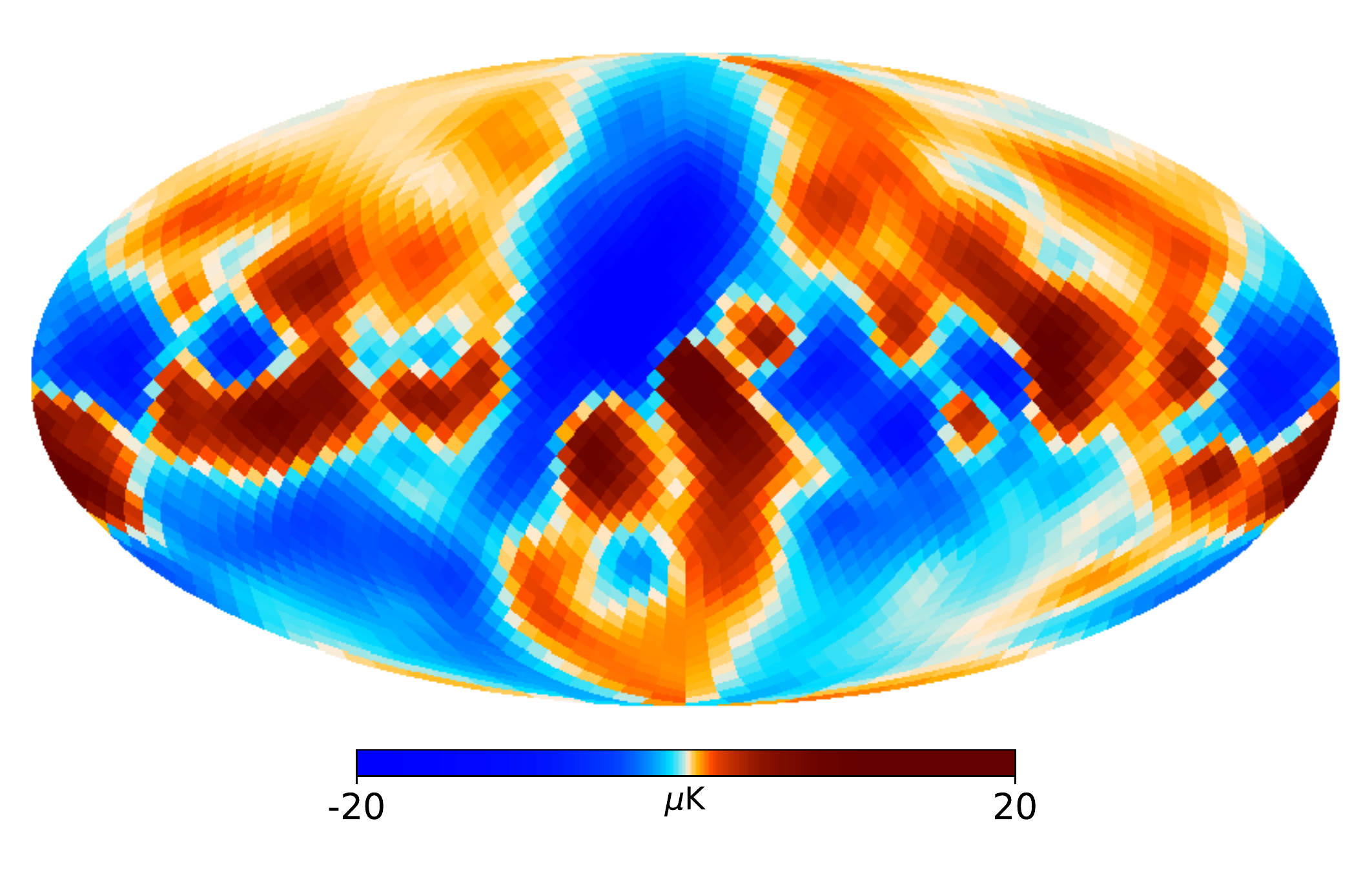}
    \caption{Stokes $Q$ (top) and $U$ (bottom) maps for the thermal dust foreground component at resolution $N_{\rm side}=16$.}
    \label{DQU1}
\end{figure}

In all cases, as expected, the foreground maps show directionality that is aligned with the Galaxy. The degraded foreground maps are shown in Figs.~\ref{SQU1} and~\ref{DQU1}.

\subsection{Sensitivity to foreground contamination}
\label{sec:foreground_contamination}

\begin{figure}
  \includegraphics[width=\columnwidth]{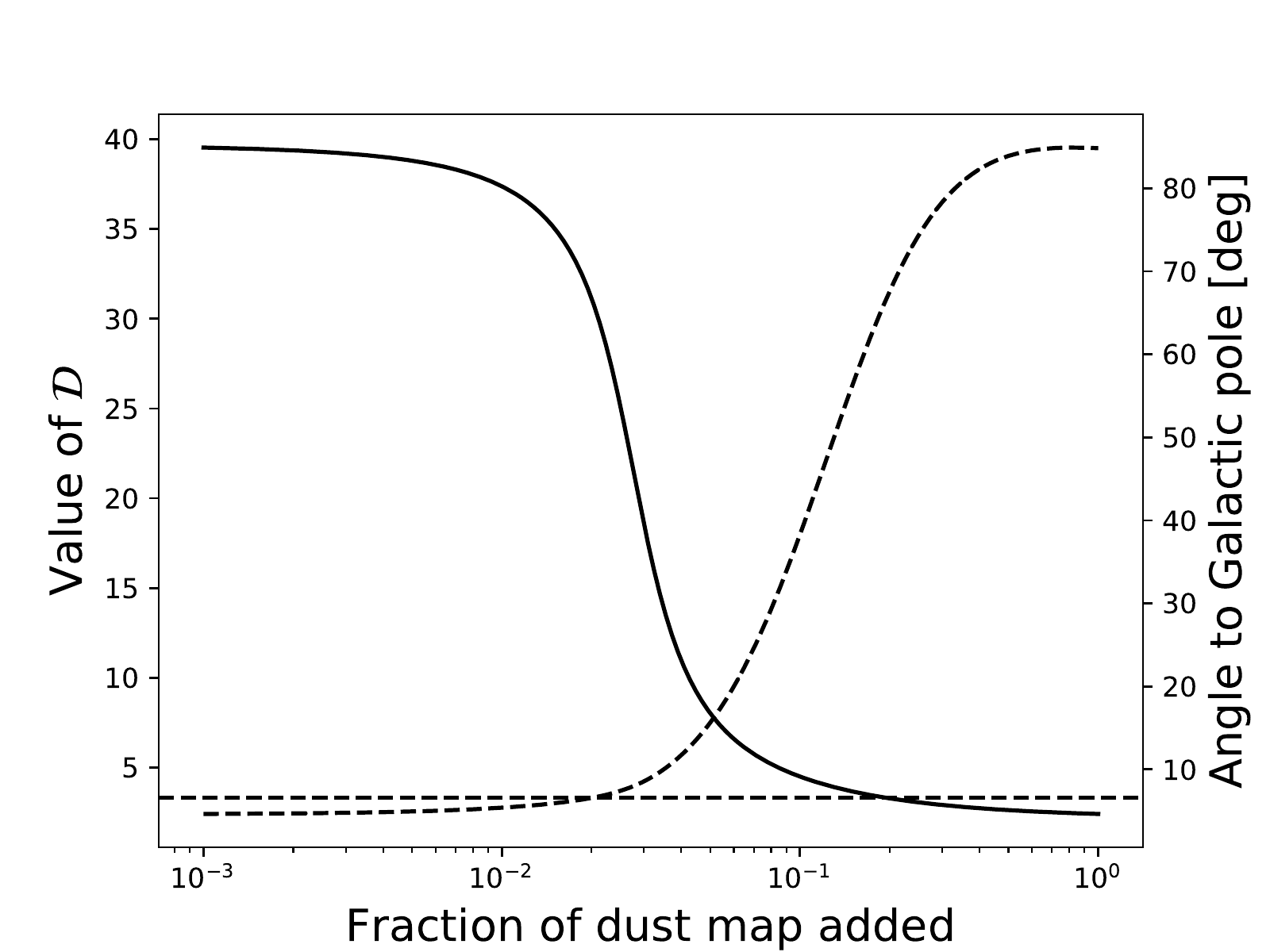} \\
  \caption{Value of $\mathcal{D}$ (dashed, left axis labels) and the angle from
  the maximal direction to the Galactic Poles (solid, right axis labels) for the
  {\tt Commander} map with a varying fraction of the polarized dust map added. The horizontal dashed line indicates the value of $\mathcal{D}$ at which the foreground will be detected with 95
  per cent confidence.}
  \label{CommD}
\end{figure}

To examine the sensitivity of $\mathcal{D}$ to foreground contamination, we analyse CMB maps with small amounts of added foregrounds. Specifically, we add a varying fraction of the polarized thermal dust and synchrotron emission maps to the \texttt{Commander} and \texttt{SMICA} maps and determine the fractional value at which the $\mathcal{D}$ statistic would detect foregrounds with 95 per cent confidence with respect to the simulations used in Section~\ref{sec:cmb}. These fractions are summarized in Table~\ref{fracs}, and in Fig.~\ref{CommD} we show the value of $\mathcal{D}$ as a function of $f$, the fraction of the polarized dust map added to the \texttt{Commander} map. To demonstrate the effect of foregrounds on directionality, Fig.~\ref{CommD} also shows the angle (from the Galactic poles) for the preferred axis as a function of $f$. We see that just 1 to 6 per cent of the foreground signal would be sufficient to see a directional signal.

\begin{table}
  \centering
  \caption{Fractions of foreground (polarized thermal dust or synchrotron)
  maps added to the CMB ({\tt Commander} and {\tt SMICA}) map
  at which the $\mathcal{D}$ statistic will detect contamination with 95
  per cent confidence.}
  \label{fracs}
  \begin{tabular}{ccc}
    \hline
    CMB Map            & Thermal dust & Synchrotron \\
    \hline
    \texttt{SMICA}     & 0.06       & 0.04\\
    \texttt{Commander} & 0.02        & 0.01\\
    \hline
  \end{tabular}
\end{table}

\subsection{Analysis of \textit{Planck} lensing potential data}
\label{sec:lensing}

The $\mathcal{D}$ statistic can also be used to analyse lensing maps by simply redefining $\mathbfit{g}_p$. We seek an alternative quantity to assess the gravitational lensing maps for directionality and a natural choice is the deflection angle, which is simply the gradient of the potential. Other choices are certainly possible, e.g., the gradient of the magnification $\kappa$ or the shear ($\gamma_+, \gamma_\times$); however, we restrict our analysis to the deflection angle due to its simple physical interpretation.

The lensing potential $\phi$ \citep[as defined by e.g.][]{lewis2006weak}, is not provided directly by the Planck Collaboration. Instead, the spherical harmonic coefficients of the estimated lensing convergence $\kappa$ are described in \citet{2016A&A...594A..15P} and provided through the PLA. Here, the convergence modes on the sky are defined by
\begin{equation}
  \kappa_{\ell m}=\frac{\ell(\ell+1)}{2}\phi_{\ell m}.
   \label{eq:25}
\end{equation}
This is a particularly useful data product because the reconstruction noise on $\kappa$ is approximately white \citep{bucher2012cmb}.

In order to obtain $\phi$, $\kappa$ must be divided by $\ell(\ell+1)/2$. After doing so, $\mathbfit{g}_p$ can be defined as the deflection angle, $\mathbf{\alpha}$ on the sky:
\begin{equation}
  \mathbfit{g}_p \equiv \mathbf{\alpha} = \nabla \phi.
  \label{eq:26}
\end{equation}
The {\tt HEALPix} function \texttt{alm2map\_der1} is used to obtain $\mathbf{\alpha}$. We do so at the resolution $N_{\rm side}=16$, which effectively corresponds to a multipole range with $\ell_{\rm max} = 64$. The mask required is provided alongside $\kappa$ in the PLA. For the simulated maps, the PLA has provided 100 simulated spherical harmonic coefficients of $\kappa$, which are processed as described above to obtain the lensing potential.

\begin{figure}
  \includegraphics[width=\columnwidth]{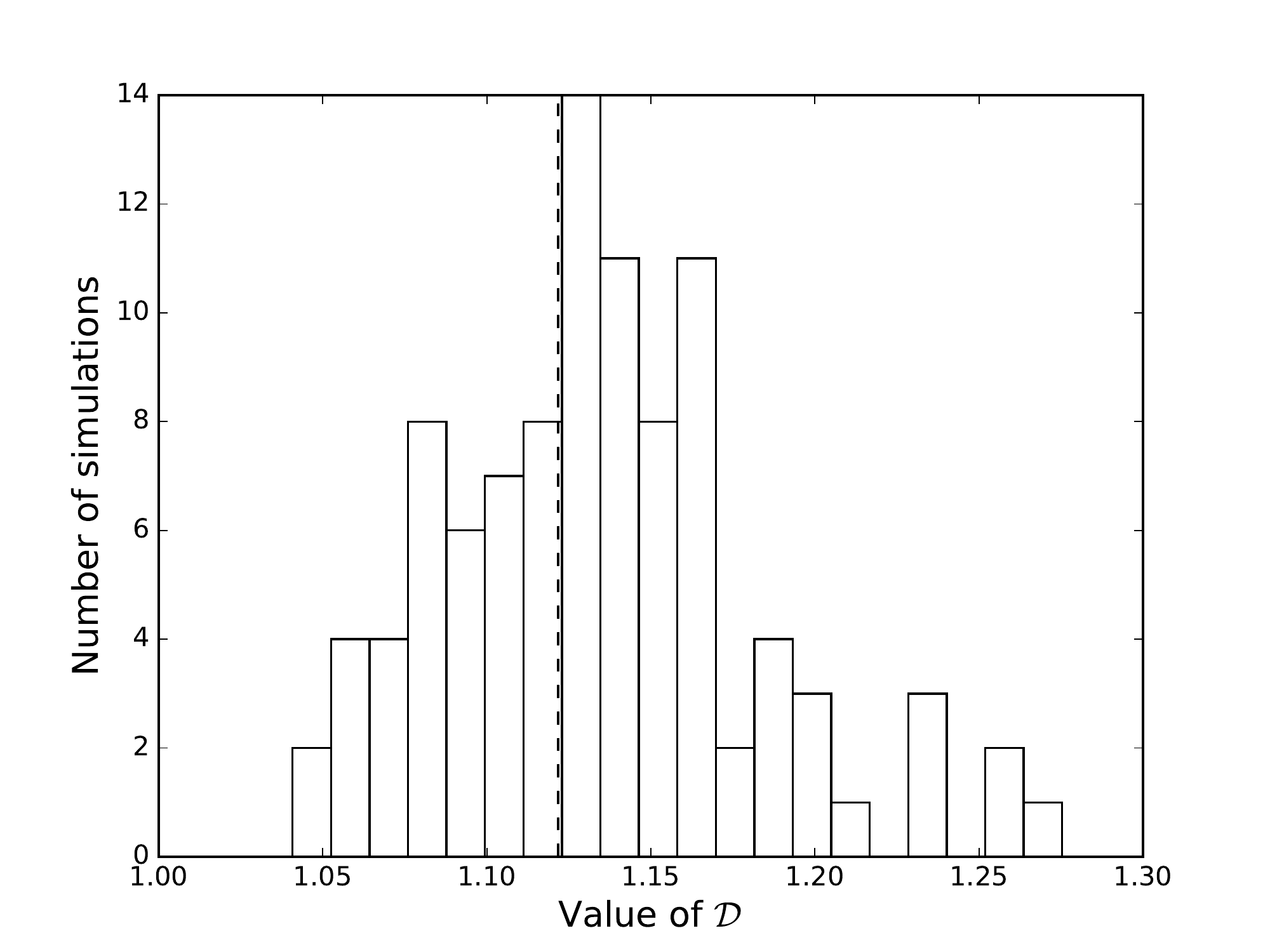}
    \caption{Directionality histogram for lensing-potential simulations. The dashed line, shown at $\mathcal{D} = 1.12$, is the value of $\mathcal{D}$ calculated using the minimum-variance lensing-potential data. Clearly the data are consistent with the simulations.  The map has been masked with the lensing mask provided by the Planck Collaboration and degraded to $N_{\rm side}=16$.}
    \label{fig:6}
\end{figure}

We may now proceed exactly as before to calculate the $\mathcal{D}$ statistic. Figure~\ref{fig:6} displays the $\mathcal{D}$ statistic value for the data, along with the distribution for the simulations. There is no sign of significant directionality in the {\it Planck\/} lensing data at large angular scales.

\section{Conclusions}
\label{sec:conclusions}

We have used the $\mathcal{D}$ statistic, introduced by \citet{2000MNRAS.313..331B}, to analyse \Planck\ polarization maps. Assessing the frequency maps, we calculated the significance $\Delta \mathcal{D}$ using a  mask that gradually increased in thickness. We found that the value of $\mathcal{D}$ lies well beyond the distribution for foreground-free simulations until the mask used is large enough to remove the Galactic foreground (as well as most of the sky).

When analysing the CMB maps, we found no excess directionality. This leads us to conclude that there is no evidence of residual foreground contamination in the \texttt{NILC}, \texttt{SMICA}, \texttt{Commander}, and \texttt{SEVEM} maps.

To examine the sensitivity of $\mathcal{D}$ to residual Galactic contamination, we tested the fractions at which foreground contamination will be detected with 95 per cent confidence. Our tests indicate that the $\mathcal{D}$ statistic is effective in detecting foreground contamination at the per cent level.

For the \Planck\ lensing potential data, we demonstrated how the $\mathcal{D}$ statistic can be used to assess directionality by taking the gradient of the map. When compared to the simulations, the minimum-variance lensing-potential map does not show any sign of directionality.

The $\mathcal{D}$ statistic is a useful tool for the purpose of detecting residual foreground and systematic effects or assessing the directionality of a map in general. It is a simple statistic that is easily computable and hence is appropriate to have as part of any tool-kit for investigating the statistical isotropy of maps of the sky.

\section*{Acknowledgements}

This research was supported by the Natural Sciences and Engineering Research Council of Canada. EFB is supported by US National Science Foundation award AST-1410133.



\bibliographystyle{mnras}
\bibliography{Dstatistic} 

\begin{thebibliography}{}
\makeatletter
\relax
\def\mn@urlcharsother{\let\do\@makeother \do\$\do\&\do\#\do\^\do\_\do\%\do\~}
\def\mn@doi{\begingroup\mn@urlcharsother \@ifnextchar [ {\mn@doi@}
  {\mn@doi@[]}}
\def\mn@doi@[#1]#2{\def\@tempa{#1}\ifx\@tempa\@empty \href
  {http://dx.doi.org/#2} {doi:#2}\else \href {http://dx.doi.org/#2} {#1}\fi
  \endgroup}
\def\mn@eprint#1#2{\mn@eprint@#1:#2::\@nil}
\def\mn@eprint@arXiv#1{\href {http://arxiv.org/abs/#1} {{\tt arXiv:#1}}}
\def\mn@eprint@dblp#1{\href {http://dblp.uni-trier.de/rec/bibtex/#1.xml}
  {dblp:#1}}
\def\mn@eprint@#1:#2:#3:#4\@nil{\def\@tempa {#1}\def\@tempb {#2}\def\@tempc
  {#3}\ifx \@tempc \@empty \let \@tempc \@tempb \let \@tempb \@tempa \fi \ifx
  \@tempb \@empty \def\@tempb {arXiv}\fi \@ifundefined
  {mn@eprint@\@tempb}{\@tempb:\@tempc}{\expandafter \expandafter \csname
  mn@eprint@\@tempb\endcsname \expandafter{\@tempc}}}

\bibitem[\protect\citeauthoryear{Bucher, Carvalho, Moodley  \&
  Remazeilles}{Bucher et~al.}{2012}]{bucher2012cmb}
Bucher M.,  Carvalho C.~S.,  Moodley K.,   Remazeilles M.,  2012, PRD, 85,
  043016

\bibitem[\protect\citeauthoryear{{Bunn} \& {Scott}}{{Bunn} \&
  {Scott}}{2000}]{2000MNRAS.313..331B}
{Bunn} E.~F.,  {Scott} D.,  2000, \mn@doi [MNRAS]
  {10.1046/j.1365-8711.2000.03212.x}, \href
  {http://adsabs.harvard.edu/abs/2000MNRAS.313..331B} {313, 331}

\bibitem[\protect\citeauthoryear{{Davis} \& {Greenstein}}{{Davis} \&
  {Greenstein}}{1951}]{1951ApJ...114..206D}
{Davis} Jr. L.,  {Greenstein} J.~L.,  1951, \mn@doi [ApJ] {10.1086/145464},
  \href {http://adsabs.harvard.edu/abs/1951ApJ...114..206D} {114, 206}

\bibitem[\protect\citeauthoryear{{Ellis} \& {Schreiber}}{{Ellis} \&
  {Schreiber}}{1986}]{1986PhLA..115...97E}
{Ellis} G.~F.~R.,  {Schreiber} G.,  1986, \mn@doi [Phys.\ Lett.\ A]
  {10.1016/0375-9601(86)90032-0}, \href
  {http://adsabs.harvard.edu/abs/1986PhLA..115...97E} {115, 97}

\bibitem[\protect\citeauthoryear{{G{\'o}rski}, {Hivon}, {Banday}, {Wandelt},
  {Hansen}, {Reinecke}  \& {Bartelmann}}{{G{\'o}rski}
  et~al.}{2005}]{2005ApJ...622..759G}
{G{\'o}rski} K.~M.,  {Hivon} E.,  {Banday} A.~J.,  {Wandelt} B.~D.,  {Hansen}
  F.~K.,  {Reinecke} M.,   {Bartelmann} M.,  2005, \mn@doi [ApJ]
  {10.1086/427976}, \href {http://adsabs.harvard.edu/abs/2005ApJ...622..759G}
  {622, 759}

\bibitem[\protect\citeauthoryear{{Hanson}, {Scott}  \& {Bunn}}{{Hanson}
  et~al.}{2007}]{2007MNRAS.381....2H}
{Hanson} D.,  {Scott} D.,   {Bunn} E.~F.,  2007, \mn@doi [MNRAS]
  {10.1111/j.1365-2966.2007.12180.x}, \href
  {http://adsabs.harvard.edu/abs/2007MNRAS.381....2H} {381, 2}

\bibitem[\protect\citeauthoryear{{Hu} \& {White}}{{Hu} \&
  {White}}{1997}]{1997NewA....2..323H}
{Hu} W.,  {White} M.,  1997, \mn@doi [NewA] {10.1016/S1384-1076(97)00022-5},
  \href {http://adsabs.harvard.edu/abs/1997NewA....2..323H} {2, 323}

\bibitem[\protect\citeauthoryear{{Kamionkowski} \& {Kosowsky}}{{Kamionkowski}
  \& {Kosowsky}}{1998}]{1998PhRvD..57..685K}
{Kamionkowski} M.,  {Kosowsky} A.,  1998, \mn@doi [PRD]
  {10.1103/PhysRevD.57.685}, \href
  {http://adsabs.harvard.edu/abs/1998PhRvD..57..685K} {57, 685}

\bibitem[\protect\citeauthoryear{{Kamionkowski}, {Kosowsky}  \&
  {Stebbins}}{{Kamionkowski} et~al.}{1997}]{1997PhRvD..55.7368K}
{Kamionkowski} M.,  {Kosowsky} A.,   {Stebbins} A.,  1997, \mn@doi [PRD]
  {10.1103/PhysRevD.55.7368}, \href
  {http://adsabs.harvard.edu/abs/1997PhRvD..55.7368K} {55, 7368}

\bibitem[\protect\citeauthoryear{Lewis \& Challinor}{Lewis \&
  Challinor}{2006}]{lewis2006weak}
Lewis A.,  Challinor A.,  2006, Phys.\ Rep., 429, 1

\bibitem[\protect\citeauthoryear{{Page} et~al.,}{{Page}
  et~al.}{2007}]{2007ApJS..170..335P}
{Page} L.,  et~al., 2007, \mn@doi [ApJS] {10.1086/513699}, \href
  {http://adsabs.harvard.edu/abs/2007ApJS..170..335P} {170, 335}

\bibitem[\protect\citeauthoryear{{Planck Collaboration}}{{Planck
  Collaboration}}{2016}]{2016planckXXV}
{Planck Collaboration} 2016, \mn@doi [\aap] {10.1051/0004-6361/201526803}, 594,
  A25

\bibitem[\protect\citeauthoryear{{Planck Collaboration}}{{Planck
  Collaboration}}{2018}]{2018planckXII}
{Planck Collaboration} 2018, arXiv e-prints

\bibitem[\protect\citeauthoryear{{Planck Collaboration I}}{{Planck
  Collaboration I}}{2014}]{2014A&A...571A...1P}
{Planck Collaboration I} 2014, \mn@doi [A\&A] {10.1051/0004-6361/201321529},
  \href {http://adsabs.harvard.edu/abs/2014A%26A...571A...1P} {571, A1}

\bibitem[\protect\citeauthoryear{{Planck Collaboration II}}{{Planck
  Collaboration II}}{2018}]{2018planckII}
{Planck Collaboration II} 2018, arXiv preprint arXiv:1807.06206

\bibitem[\protect\citeauthoryear{{Planck Collaboration III}}{{Planck
  Collaboration III}}{2018}]{2018planckIII}
{Planck Collaboration III} 2018, arXiv preprint arXiv:1807.06207

\bibitem[\protect\citeauthoryear{{Planck Collaboration IV}}{{Planck
  Collaboration IV}}{2018}]{2018planckIV}
{Planck Collaboration IV} 2018, arXiv preprint arXiv:1807.06208

\bibitem[\protect\citeauthoryear{{Planck Collaboration IX}}{{Planck
  Collaboration IX}}{2016}]{2016A&A...594A...9P}
{Planck Collaboration IX} 2016, \mn@doi [A\&A] {10.1051/0004-6361/201525936},
  \href {http://adsabs.harvard.edu/abs/2016A%26A...594A...9P} {594, A9}

\bibitem[\protect\citeauthoryear{{Planck Collaboration VIII}}{{Planck
  Collaboration VIII}}{2016}]{2016A&A...594A...8P}
{Planck Collaboration VIII} 2016, \mn@doi [A\&A] {10.1051/0004-6361/201525820},
  \href {http://adsabs.harvard.edu/abs/2016A%26A...594A...8P} {594, A8}

\bibitem[\protect\citeauthoryear{{Planck Collaboration X}}{{Planck
  Collaboration X}}{2016}]{2016A&A...594A..10P}
{Planck Collaboration X} 2016, \mn@doi [A\&A] {10.1051/0004-6361/201525967},
  \href {http://adsabs.harvard.edu/abs/2016A%26A...594A..10P} {594, A10}

\bibitem[\protect\citeauthoryear{{Planck Collaboration XIII}}{{Planck
  Collaboration XIII}}{2016}]{2016A&A...594A..13P}
{Planck Collaboration XIII} 2016, \mn@doi [A\&A] {10.1051/0004-6361/201525830},
  \href {http://adsabs.harvard.edu/abs/2016A%26A...594A..13P} {594, A13}

\bibitem[\protect\citeauthoryear{{Planck Collaboration XV}}{{Planck
  Collaboration XV}}{2016}]{2016A&A...594A..15P}
{Planck Collaboration XV} 2016, \mn@doi [A\&A] {10.1051/0004-6361/201525941},
  \href {http://adsabs.harvard.edu/abs/2016A%26A...594A..15P} {594, A15}

\bibitem[\protect\citeauthoryear{{Planck Collaboration XVI}}{{Planck
  Collaboration XVI}}{2016}]{collaboration2015planck}
{Planck Collaboration XVI} 2016, \mn@doi [A\&A] {10.1051/0004-6361/201526681},
  594, A16

\bibitem[\protect\citeauthoryear{{Seljak} \& {Zaldarriaga}}{{Seljak} \&
  {Zaldarriaga}}{1997}]{1997PhRvL..78.2054S}
{Seljak} U.,  {Zaldarriaga} M.,  1997, \mn@doi [PRL]
  {10.1103/PhysRevLett.78.2054}, \href
  {http://adsabs.harvard.edu/abs/1997PhRvL..78.2054S} {78, 2054}

\bibitem[\protect\citeauthoryear{{Stevens}, {Scott}  \& {Silk}}{{Stevens}
  et~al.}{1993}]{1993PhRvL..71...20S}
{Stevens} D.,  {Scott} D.,   {Silk} J.,  1993, \mn@doi [PRL]
  {10.1103/PhysRevLett.71.20}, \href
  {http://adsabs.harvard.edu/abs/1993PhRvL..71...20S} {71, 20}

\bibitem[\protect\citeauthoryear{{de Oliveira-Costa}, {Smoot}  \&
  {Starobinsky}}{{de Oliveira-Costa} et~al.}{1996}]{1996ApJ...468..457D}
{de Oliveira-Costa} A.,  {Smoot} G.~F.,   {Starobinsky} A.~A.,  1996, \mn@doi
  [ApJ] {10.1086/177706}, \href
  {http://adsabs.harvard.edu/abs/1996ApJ...468..457D} {468, 457}

\makeatother
\end{thebibliography}



\appendix

\section{Distinguishing Between \textit{E} modes and $\textit{B}$ modes}
\label{sec:AppA}

An interesting observation we made while conducting this research is that the $\mathcal{D}$ statistic is capable of distinguishing between \textit{E} modes and $\textit{B}$ modes. Polarization patterns are decomposed into \textit{E} modes (a part that comes from a divergence), and $\textit{B}$ modes (a part that comes from a curl). The divergence pattern will tend to have polarization directions that are more aligned with each other than a pattern coming from a curl, and so it is expected that an \textit{E} mode will have a higher $\mathcal{D}$ statistic than a $\textit{B}$ mode.

To illustrate this we calculate the $\mathcal{D}$ statistic, by hand, for the $\ell\,{=}\,2$, $m\,{=}\,0$ quadrupole. For this demonstration we calculate $\mathcal{D}$ for $a_{2,0}^E = -1$  and then again for $a_{2,0}^B = -1$. The $\mathcal{D}$ statistic is calculated at  $N_{\rm side}=1$, meaning we only consider 12 polarization pseudo-vectors on the sphere. The positions of the pseudo-vectors, as well as the $\textit{Q}$ and $\textit{U}$ values for both the $\textit{E}$-mode and \textit{B}-mode quadrupoles, are specified in Table~\ref{tab:example_table}.

\begin{table}
  \centering
  \caption{Polarization pseudo-vector positions and stokes \textit{Q} and \textit{U} values in the \textit{E}-mode quadrupole and \textit{B}-mode quadrupole for an $N_{\rm side}=1$ sky-map. These values are used to calculate the $\mathcal{D}$ value by hand for each of the two cases}
  \label{tab:example_table}
  \begin{tabular}{ccccccc}
    \hline
    Vector No. & $\theta$ [rad]& $\phi$ [Rad]& $Q_{E}$ & $U_{E}$ & $Q_{B}$ & $U_{B}$\\
    \hline
    1 & 0.841 & 0.786 & 0.215 & 0 & 0 & 0.215\\
    2 & 0.841 & 2.356 & 0.215 & 0 & 0 & 0.215\\
    3 & 0.841 & 3.927 & 0.215 & 0 & 0 & 0.215\\
    4 & 0.841 & 5.498 & 0.215 & 0 & 0 & 0.215\\
    5 & 1.571 & 0.000 & 0.386 & 0 & 0 & 0.386\\
    6 & 1.571 & 1.571 & 0.386 & 0 & 0 & 0.386\\
    7 & 1.571 & 3.142 & 0.386 & 0 & 0 & 0.386\\
    8 & 1.571 & 4.712 & 0.386 & 0 & 0 & 0.386\\
    9 & 2.301 & 0.786 & 0.215 & 0 & 0 & 0.215\\
    10 & 2.301 & 2.356 & 0.215 & 0 & 0 & 0.215\\
    11 & 2.301 & 3.927 & 0.215 & 0 & 0 & 0.215\\
    12 & 2.301 & 5.498 & 0.215 & 0 & 0 & 0.215\\
    \hline
  \end{tabular}
\end{table}

Given \textit{Q} and \textit{U}, we can calculate $\gamma$ for each vector, as described in Section~\ref{sec:dstat}. For our simple $\textit{E}$-mode example we have $\gamma = 0$ for all vectors. Recall that, since the $\mathcal{D}$-statistic is quadratic, we can treat psuedo-vectors as vectors in the northern half of the tangent plane. For the $\textit{B}$-mode example, we have $\gamma = \pi/4$ for all of the vectors. It is clear that vectors with $\gamma = 0$ align well with the \textit{z}-axis and poorly with the \textit{xy}-plane; since the $\mathcal{D}$ statistic is a ratio of the maximum and minimum values of $f(\hat{\mathbfit{n}})$, we expect that this will result in a large $\mathcal{D}$ statistic compared to vectors that have $\gamma = \pi/4$. Using $\gamma$, $\theta$, $\phi$, \textit{Q} and \textit{U}, we can calculate the vectors as
\begin{align}
\begin{split}
    \mathbfit{g}_p &= \sqrt{Q^2+U^2} \begin{bmatrix}
           (-\cos{\gamma}\cos{\phi}\cos{\theta}) + (\sin{\gamma}\sin{\phi})\\
           (-\cos{\gamma}\sin{\phi}\cos{\theta}) - (\sin{\gamma}\cos{\phi})\\
           \cos{\gamma}\sin{\theta}
         \end{bmatrix}.
    \label{eq:27}
\end{split}
\end{align}

Now we can determine the $\mathcal{D}$ statistic for both situations by maximizing and minimizing $f(\hat{\mathbfit{n}})$, as defined in equation~(\ref{f}). Since there is no masking or noise involved, we assume that the weights are all 1. Following the argument presented in equations~(\ref{Eigen1}) and ~(\ref{Eigen2}), we reduce this to an eigenvalue problem for matrix $\mathbfss{A}$, as defined in equation~(\ref{Eigen1}). We start with the $\textit{E}$-mode quadrupole, for which
\begin{align}
\begin{split}
    \mathbfss{A} &= \begin{bmatrix} \phantom{-}8.22\times10^{-2} & -1.15\times10^{-8} & -1.83\times10^{-8} \\
                                -1.15\times10^{-8} & \phantom{-}8.23\times10^{-2} & -1.09\times10^{-8}\\
                                -1.83\times10^{-8} & -1.09\times10^{-8} & \phantom{-}8.01\times10^{-1}\end{bmatrix}.
    \label{eq:28}
\end{split}
\end{align}
The maximum and minimum eigenvalues for this matrix are 0.801 and 0.0822, and thus the $\mathcal{D}$ statistic is 9.74; the maximum eigenvector is $\big[-2.55\times10^{-8},-1.52\times10^{-8},1.00\big]$, which points toward the \textit{z}-axis, as expected. Repeating the procedure for the $\textit{B}$-mode quadrupole we find maximum and minimum eigenvalues of 0.401 and 0.283; thus the $\mathcal{D}$ statistic is 1.42 and the maximum eigenvector is $\big[-5.11\times10^{-5},4.14\times10^{-4},1.00\big]$.

\begin{figure}
  \includegraphics[width=\columnwidth]{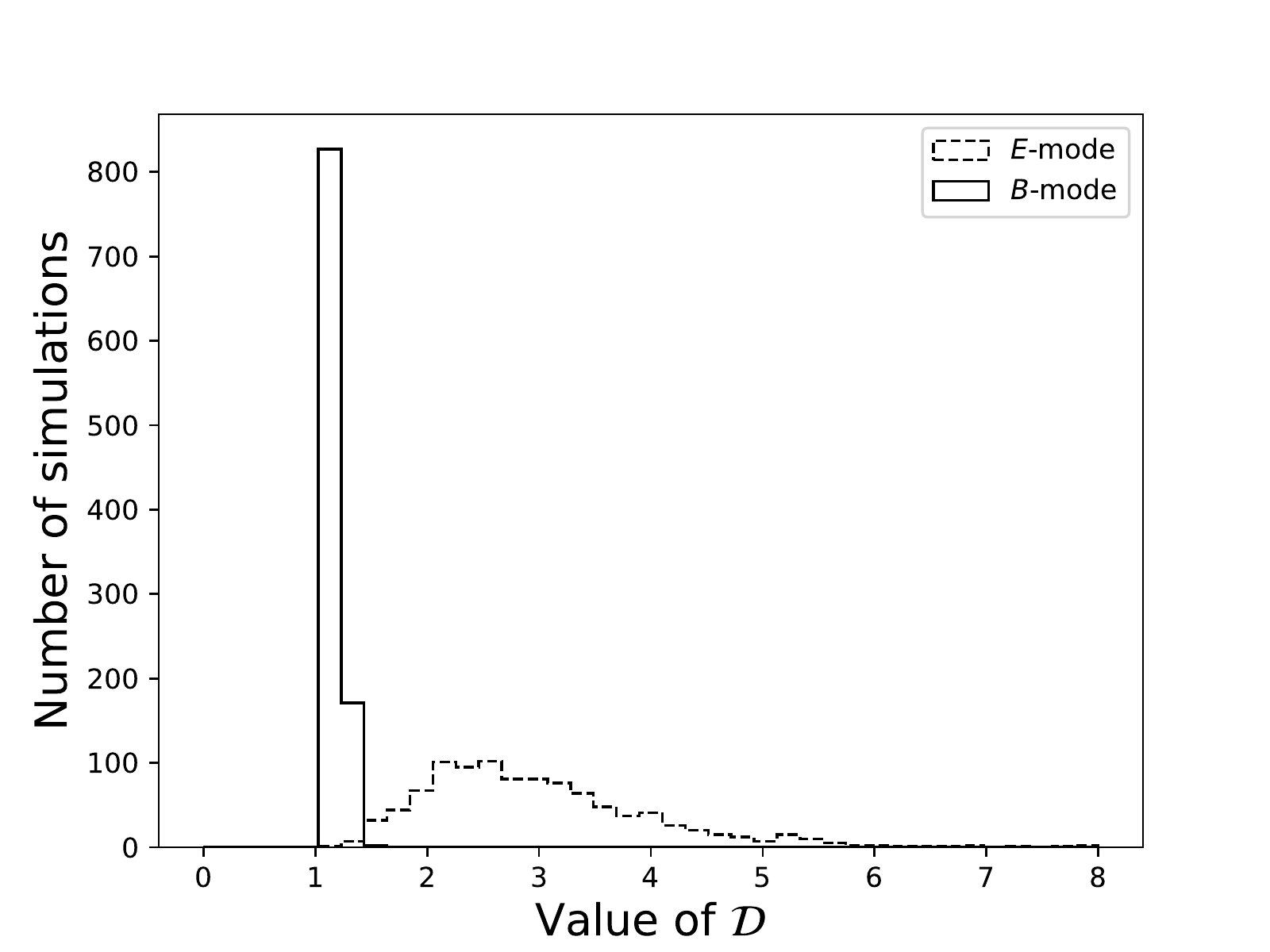}
    \caption{$\mathcal{D}$ statistic distributions for $\textit{E}$-mode and $\textit{B}$-mode polarization patterns. The $\textit{E}$-mode pattern is generated using the theory $C_\ell^{EE}$ power spectrum for the best-fit $\Lambda$CDM model provided by the Planck Collaboration. The $\textit{B}$-mode polarization pattern is generated by substituting the same $C_\ell^{EE}$ power spectrum values into $C_\ell^{BB}$ and treating this as a pure $\textit{B}$-mode power spectrum.}
    \label{fig:A2}
\end{figure}

\begin{figure}
  \begin{tabular}{@{}c@{}}
    \includegraphics[width=\columnwidth]{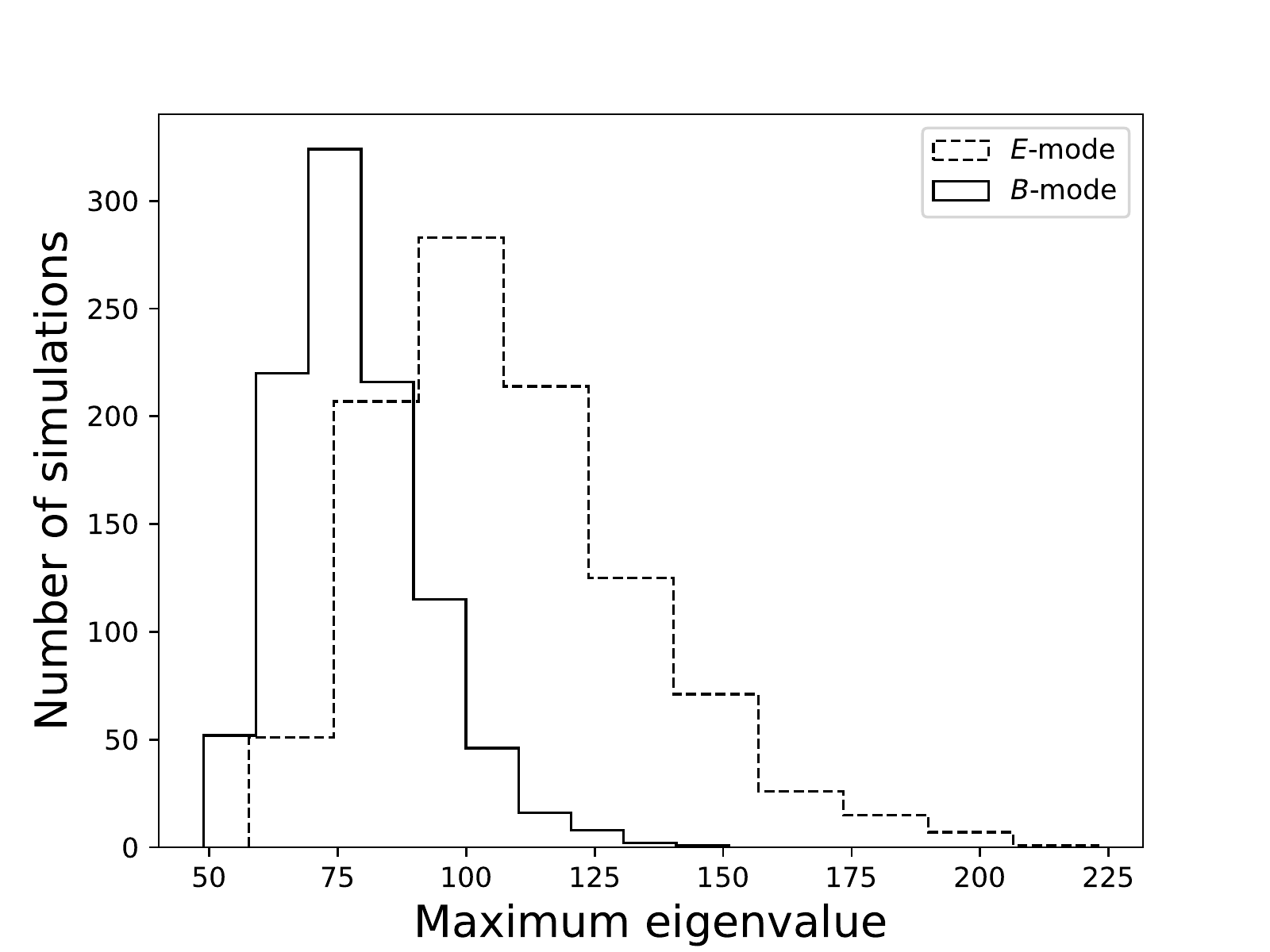} \\
    \includegraphics[width=\columnwidth]{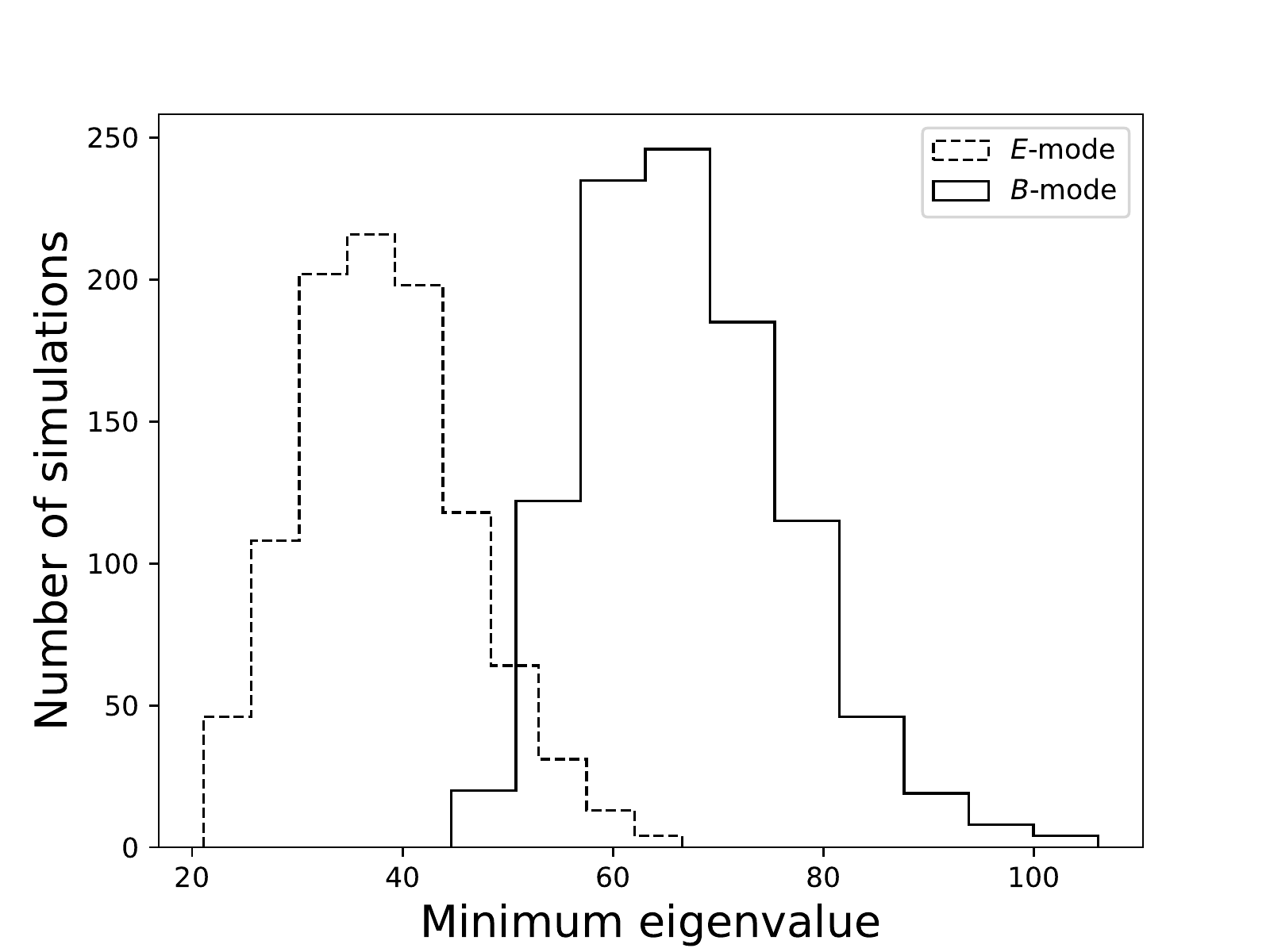} \\
  \end{tabular}
  \caption{Distributions of maximum (top panel) and minimum (bottom panel) eigenvalues of $\mathbfss{A}$ for $\textit{E}$-mode and $\textit{B}$-mode polarization patterns. The $\textit{E}$-mode polarization pattern is generated using the theory $C_\ell^{EE}$ power spectrum for the best-fit $\Lambda$CDM model provided by the Planck Collaboration. The $\textit{B}$-mode polarization pattern is generated by substituting the same $C_\ell^{EE}$ power spectrum values into $C_\ell^{BB}$ and treating it as a pure $\textit{B}$-mode power spectrum.}
  \label{fig:A3}
\end{figure}

To further test how the $\mathcal{D}$ statistic distinguishes between $\textit{E}$ modes and $\textit{B}$ modes, we analyse CMB \textit{E}-mode simulations. More specifically, we generate simulations with only an $EE$ power spectrum consistent with that obtained by the Planck Collaboration. We then analyse $\textit{B}$-mode simulations, this time using only $BB$ power, where the values of $C_\ell^{BB}$ are replaced with $C_\ell^{EE}$. Since our purpose here is to provide a simple illustration of the difference in the way the statistic treats $\textit{E}$ modes and $\textit{B}$ modes, we ignore the temperature signal and the associated TE correlations which would be necessary in a full analysis. The result is shown in Fig.~\ref{fig:A2}. Figure~\ref{fig:A3} shows the distributions of the minimum and maximum eigenvalues for both sets of simulations, demonstrating that they are indeed quite different.

\section{Detecting rotations in the CMB}
\label{sec:AppB}

\begin{figure}
  \includegraphics[width=\columnwidth]{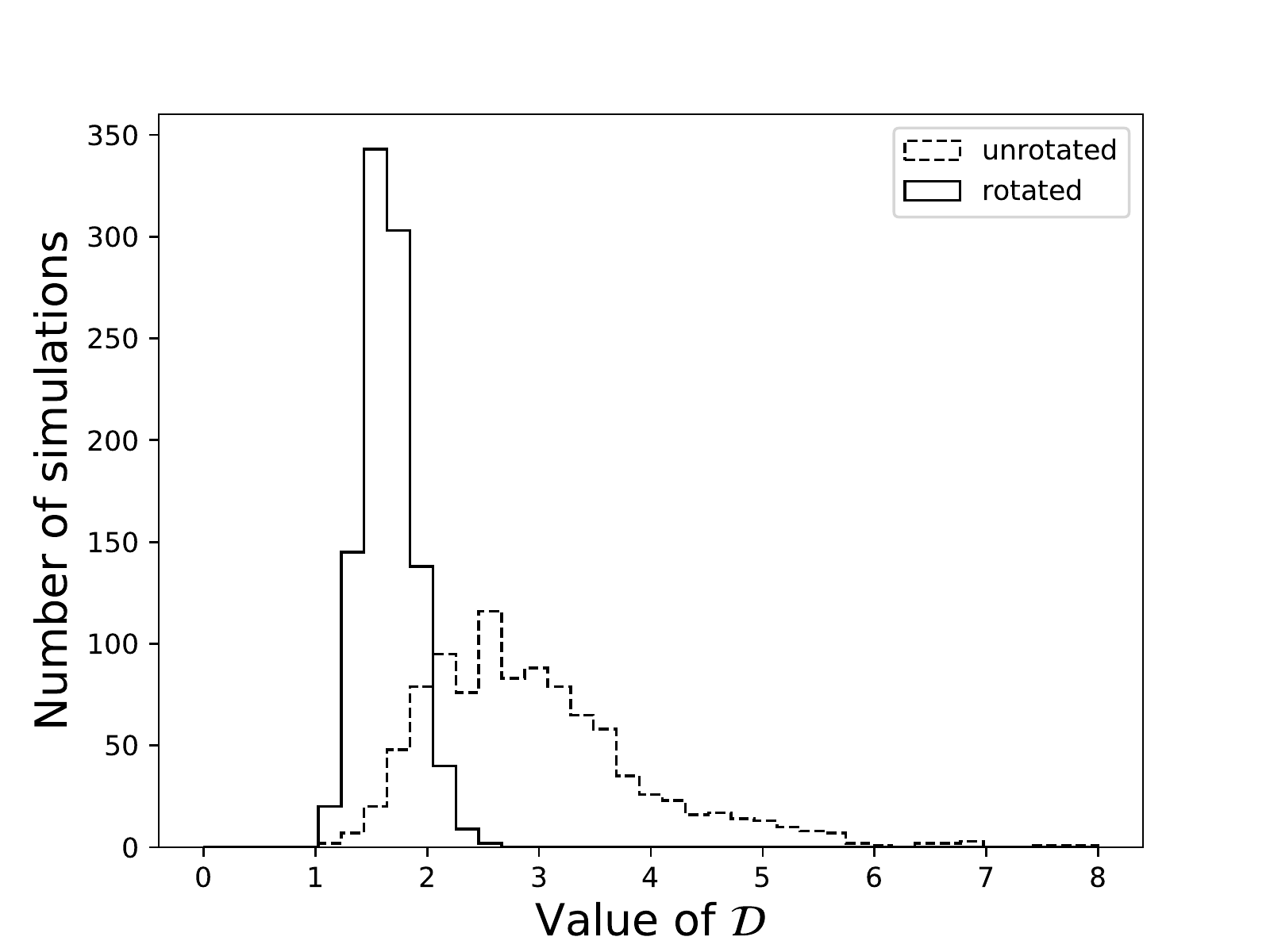}
    \caption{$\mathcal{D}$ statistic calculated for CMB simulations (generated using the best $\Lambda$CDM theory CMB power spectra) and once again for the same set of simulations rotated by $\pi /3$.}
    \label{fig:A5}
\end{figure}

\begin{figure}
  \includegraphics[width=\columnwidth]{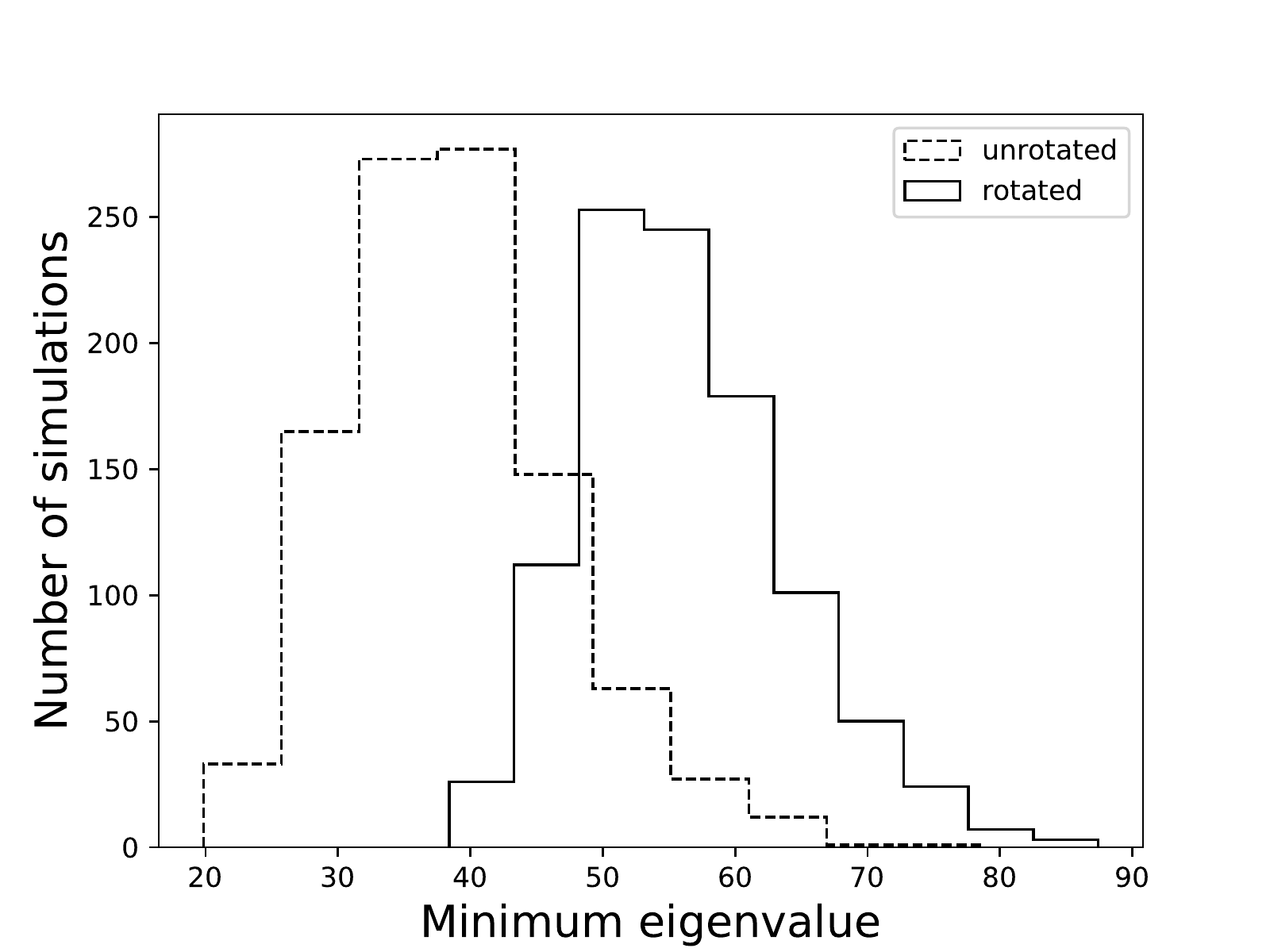}
    \caption{Minimum eigenvalue of matrix $\mathbfss{A}$ calculated for CMB simulations (generated using the best $\Lambda$CDM theory CMB power spectra) and once again for the same set of simulations rotated by $\pi /3$.}
    \label{fig:A6}
\end{figure}

Since the $\mathcal{D}$ statistic is capable of distinguishing between $\textit{E}$ modes and $\textit{B}$ modes, it must also be sensitive to rotations. In other words, when the $\mathcal{D}$ statistic is analysed for a set of \textit{Q} and \textit{U} data, the distribution of the $\mathcal{D}$ statistic is distinguishable from the distribution obtained by analysing the same set data rotated by $e^{2i\alpha}$. Figure~\ref{fig:A5} demonstrates this for an $\alpha = \pi~/3$ rotation (this is an arbitrary illustrative example); we see that the distributions of the lower eigenvalue of $\mathbfss{A}$ for rotated and unrotated simulations have little overlap. Therefore it is possible to use the lower eigenvalue of $\mathbfss{A}$ as a quantity that can distinguish whether a given simulation  belongs to the rotated or unrotated data set, as shown in Fig.~\ref{fig:A6}.

\bsp	
\label{lastpage}
\end{document}